\documentclass[JHEP,12pt]{article}

\usepackage{booktabs}

\usepackage{jheppub}
\usepackage{amssymb,amsmath,amsthm}
\usepackage[shortlabels]{enumitem}
\usepackage{braket}
\usepackage{breqn,hyperref}
\usepackage{comment} 
\usepackage{subcaption}
\usepackage{graphicx, xcolor, varwidth}
\usepackage{color}
\usepackage{float}

\newtheorem{thm}{Theorem}
\theoremstyle{definition}
\newtheorem{defn}[thm]{Definition}

\theoremstyle{remark}
\newtheorem{rem}[thm]{Remark}

\renewcommand{\S}{S_{\rm gen}}
\newcommand{\emax}{e_{\rm max}}
\newcommand{\emin}{e_{\rm min}}

\newcommand{\V}{\mathcal{V}}
\newcommand{\E}{\mathcal{E}}

\definecolor{bcolor}{rgb}{0.1,0,1}

\definecolor{bcolor}{rgb}{0.1,0,1}

\renewcommand{\(}{\left(}
\renewcommand{\)}{\right)}
\renewcommand{\[}{\left[}
\renewcommand{\]}{\right]}

\let\pa=\partial

\newcommand{\Eqref}[1]{Eq.~\eqref{#1}}
\newcommand{\secref}[1]{Sec.~\ref{#1}}
\newcommand{\figref}[1]{Fig.~\ref{#1}}

\title{Hollow-grams: Generalized Entanglement Wedges from the Gravitational Path Integral}
\author{Sami Kaya, Pratik Rath and Kyle Ritchie}

\affiliation{Leinweber Institute for Theoretical Physics and Department of Physics,\\
University of California, Berkeley, California 94720, U.S.A.
} 

\emailAdd{samikaya@berkeley.edu}
\emailAdd{pratik\_rath@berkeley.edu}
\emailAdd{kyle\_ritchie@berkeley.edu}
\makeatletter
\gdef\@fpheader{\mbox{}}
\makeatother

\abstract{
Recently, Bousso and Penington (BP) made a proposal for the entanglement wedge associated to a gravitating bulk region. In this paper, we derive this proposal in time-reflection symmetric settings using the gravitational path integral. To do this, we exploit the connection between random tensor networks (RTNs) and fixed-geometry states in gravity. We define the entropy of a bulk region in an RTN by removing tensors in that region and computing the entropy of the open legs thus generated in the ``hollowed" RTN. We thus derive the BP proposal for RTNs and hence, also for fixed-geometry states in gravity. By then expressing a general holographic state as a superposition over fixed-geometry states and using a diagonal approximation, we provide a general gravitational path integral derivation of the BP proposal. We demonstrate that the saddles computing the R\'enyi entropy $S_n$ depend on how the bulk region is gauge-invariantly specified. Nevertheless, we show that the BP proposal is universally reproduced in the $n\to1$ limit.
}

\begin{document}
\maketitle
\section{Introduction}

In recent years, the gravitational path integral has emerged as a central tool in our understanding of quantum gravity, providing a framework for addressing profound questions about the nature of spacetime and quantum mechanics. Although it has not been derived as a controlled approximation to an underlying theory,\footnote{An exception is two spacetime dimensions where Jackiw-Teitelboim (JT) gravity and its variants have been defined using the gravitational path integral, and correspond to ensemble averages over boundary theories \cite{Saad:2019lba,Stanford:2019vob,Maxfield:2020ale,Witten:2020wvy}.} it has nevertheless been successful in computing various coarse-grained features expected of quantum gravitational systems such as the spectral form factor at late times \cite{Saad:2018bqo}. 

One of its most striking applications has been the derivation of the Page curve \cite{Penington:2019kki,Almheiri:2019qdq}, which captures the unitary evolution of black hole evaporation. Although these calculations were performed using an auxiliary non-gravitational bath to collect Hawking radiation for mathematical convenience, the physically relevant setting is one where all regions, including the radiation, are themselves gravitating. This raises fundamental questions about how entanglement and entropy should be defined in gravitational systems and has been critiqued in Refs.~\cite{Geng:2021hlu,Raju:2021lwh}.

The challenge of extending the notion of entanglement entropy to fully gravitational systems was addressed by Bousso and Penington (BP) \cite{Bousso:2022hlz,Bousso:2023sya}, who made a proposal to identify entanglement wedges for gravitating regions.\footnote{Ref.~\cite{Dong:2020uxp} also addressed this question in a slightly different way by proposing an answer for the effective entropy of bulk quantum fields in a gravitating region. Our analysis helps clarify the difference between these two proposals.} In a time-reflection symmetric setting, the BP proposal states that the generalized entanglement wedge $E(a)$ of an input bulk region $a$ is the smallest generalized entropy region that includes $a$ and has the same conformal boundary as $a$.

The BP proposal provides a natural generalization of the Ryu-Takayanagi formula \cite{Ryu:2006bv} to the case of bulk regions and has many non-trivial properties expected of an entanglement wedge \cite{Bousso:2022hlz}. Furthermore, these generalized entanglement wedges also satisfy strong subadditivity \cite{Bousso:2022hlz}, monogamy of mutual information \cite{Bousso:2024ysg} and the rest of the inequalities of the static holographic entropy cone \cite{Bousso:2025mdp}. 

It is thus of great interest to derive the BP proposal from the gravitational path integral. This issue has been discussed in Ref.~\cite{Balasubramanian:2023dpj}, who motivated their proposal using edge modes. Since edge modes in gravity are not well understood,\footnote{E.g., in JT gravity, they have subtle issues related to the Plancherel measure \cite{Lin:2017uzr,Lin:2018xkj,Jafferis:2019wkd}.} we will instead discuss a different method motivated by tensor networks and implemented directly using the gravitational path integral. Our analysis allows us to do explicit calculations in various examples, providing a clear understanding of the gravitational path integral derivation. We also discover novel conceptual aspects of the BP proposal that we outline below.

We will derive the BP proposal in time-reflection symmetric situations by first explaining what the proposal means in a random tensor network (RTN) in \secref{sec:RTN}. RTNs, reviewed in \secref{sub:RTN}, are useful toy models for holography~\cite{Hayden:2016cfa} that have provided interesting insights into the entanglement structure of holographic states \cite{Dong:2021clv,Akers:2021pvd,Akers:2022zxr,Akers:2023fqr,Akers:2024pgq}. In \secref{sub:genEW}, we define a generalized entanglement wedge associated to a bulk region in an RTN by first deleting the tensors in that given region and generating a new ``hollowed" RTN. We then compute the entropy of the legs that were cut open in the process to define the generalized entanglement wedge. 

In the hollowed RTN, there is no qualitative difference between the open legs at the original boundary and the new open legs in the hollowed region. Thus, we can use the known technique to compute the $n$th R\'enyi entropy in an RTN using its effective description as an Ising-like model on the permutation group acting on the $n$ replica copies \cite{Hayden:2016cfa}. By doing so, we confirm that the entanglement wedge of the hollowed region matches the generalized entanglement wedge defined by the BP proposal. 

The advantage of our RTN discussion is that it provides us with a concrete Hilbert space interpretation of the BP proposal. Using this, we clarify how the BP proposal should work when multiple bulk regions are involved. In particular, in \secref{sub:ind}, we propose restricted entanglement wedges as a natural candidate for computing entropies of spacelike separated bulk regions. This leads us to a new version of the holographic inequalities, which we call restricted entropy inequalities. The unrestricted entropy inequalities previously considered required stronger independence conditions and follow from our restricted inequalities in the special case when these independence conditions are satisfied. This unifies the discussion of strong subadditivity (SSA), which only required spacelike separated subregions \cite{Bousso:2022hlz,Bousso:2023sya}, with the other holographic inequalities.

Having understood the BP proposal in RTNs, we can then translate our results to gravity by exploiting the connection between RTNs and fixed-geometry states in gravity \cite{Akers:2018fow,Dong:2018seb,Dong:2019piw,Penington:2019kki,Akers:2022zxr,Penington:2022dhr}. In \secref{sec:holEE}, we review how this works for the purpose of computing the entropy of boundary subregions in holographic states. Of course, holographic states in gravitational theories are usually prepared using a Euclidean path integral with specified asymptotic boundary conditions, and thus are quite unlike fixed-geometry states. They are more analogous to coherent states, while the fixed-geometry states are analogous to position eigenstates. Nevertheless, a semiclassical state can be expressed as a superposition over fixed-geometry states with the wavefunction computed using the gravitational path integral \cite{Marolf:2020vsi,Akers:2020pmf,Dong:2023bfy,Penington:2024jmt}. By then employing a diagonal approximation that restricts to replica-symmetric saddles, one can reproduce the standard Lewkowycz-Maldacena derivation of the Renyi entropy \cite{Lewkowycz:2013nqa,Dong:2016fnf}.\footnote{In fact, the diagonal approximation method gives the correct answer for the $n<1$ Renyi entropy while the Lewkowycz-Maldacena method leads to the wrong answer \cite{Dong_2024}.}

By employing the same idea for bulk subregions, we propose a gravitational path integral derivation of the Bousso-Penington proposal in \secref{sec:main}. In particular, for general holographic states prepared using a Euclidean path integral, the saddles we obtain include backreaction effects that change the geometry as a function of the Renyi index $n$. Our derivation thus generalizes the Lewkowycz-Maldacena computation for boundary regions \cite{Lewkowycz:2013nqa}.

While our analysis could also directly be phrased using the gravitational path integral, we find it useful to motivate the derivation using RTNs since it provides a clear formulation of what entropy it is that we are computing in the underlying theory. In particular, this gives a clear motivation for why the generalized entanglement wedge is restricted to move outward from the bulk region of interest.

An important point that we would like to emphasize is that an input to our computation is a gauge-invariant definition of the bulk region which allows us to define what we mean by the specified region in arbitrary geometries. The saddles computing the R\'enyi entropy end up depending on the precise way in which the region is specified, but the dependence drops out in the $n\to 1$ limit. Thus, we find that the generalized entanglement wedge proposed by BP is recovered independent of the details of how the bulk region is gauge-invariantly defined. 

Finally, in order to illustrate our derivation, we provide various examples in \secref{sec:eg}. In \secref{sub:JT}, we first consider the bulk region to be a finite interval in Jackiw-Teitelboim (JT) gravity. We compute the R\'enyi entropy using the gravitational path integral for different choices of gauge-invariant specification of such a region. In \secref{sub:ein}, we then do a similar computation in higher dimensional Einstein gravity for an annulus outside a black hole horizon using a minisuperspace approximation, restricting to spherically symmetric configurations.

In \secref{sec:disc}, we discuss generalizations of our work. In \secref{sub:min}, we explain how one can use the gravitational path integral to compute the min- and max-entanglement wedges for boundary subregions \cite{Akers:2020pmf,Akers:2023fqr}. The same idea can then be used for bulk regions as well using our idea of hollowing out the region. Finally, in \secref{sub:time}, we discuss the main open problem of extending our method to time-dependent situations.

\section{BP Proposal in RTNs}
\label{sec:RTN}

In this section, we review the definition of RTNs and present our proposal for generalized entanglement wedges within these models, confirming the BP proposal. Based on this analysis, we clarify the notion of independence of bulk subregions.

\subsection{Review of RTNs}
\label{sub:RTN}
Here we briefly review random tensor networks; for a more detailed exposition, we refer the reader to Ref.~\cite{Hayden:2016cfa}. A tensor network is defined by starting with a graph $G = \{ \mathcal{V}, \mathcal{E} \}$ consisting of a set of vertices $\mathcal{V}$ and edges $\mathcal{E}$. In the context of holography, this graph can be interpreted as a coarse-grained version of a smooth spatial geometry. At each vertex $x \in \mathcal{V}$, we place a rank-$k$ tensor $T_{x, \mu_1 \mu_2 \cdots \mu_k}$, where the indices $\mu_i$ correspond to the edges connected to $x$. Each edge is associated with a Hilbert space $\mathcal{H}_i$, spanned by basis vectors $\ket{\mu_i}$. The tensor $T_x$ then defines a state
\begin{equation}
	\ket{T_x} = T_{x, \mu_1 \mu_2 \cdots \mu_k} \ket{\mu_1} \ket{\mu_2} \cdots \ket{\mu_k}
\end{equation}
on the product Hilbert space $\bigotimes_i \mathcal{H}_i$, where the product runs over all the edges leaving the vertex $x$. For simplicity, all edge Hilbert spaces are taken to have the same bond dimension $\chi$, and we will be interested in the regime $\chi \gg 1$.

To define a bulk-boundary tensor network state, we contract the tensors along their shared edges to obtain:
\begin{equation}
    \left( \bigotimes_{\(xy\) \in \mathcal{E}} \bra{xy} \right) \left( \bigotimes_{x \in \mathcal{V}} \ket{T_x} \right) ~,
\end{equation}
where $\ket{xy} = \sum_{\mu} \ket{\mu}_x \ket{\mu}_y$ is a maximally entangled state defined on the Hilbert space associated with the edge $\( xy\)$ and such contracted legs are called in-plane legs. The contraction is implemented by projecting onto these maximally entangled states, with the appropriate index of tensor $T_x$ contracted with that of tensor $T_y$. The resulting state then lives in the Hilbert space of the remaining open legs, i.e., $\mathcal{H}_{\text{bulk}} \otimes \mathcal{H}_\partial$, where $\mathcal{H}_\partial$ is associated with the legs on the boundary of the tensor network, and $\mathcal{H}_{\text{bulk}}$ corresponds to the bulk dangling legs. This bulk-boundary tensor network state can also equivalently be interpreted as a map from bulk to boundary, i.e., from $\mathcal{H}_{\text{bulk}}$ to $\mathcal{H}_\partial$.

In general, the graph $G$ can be arbitrary, but for concreteness, the reader may focus on a model inspired by AdS/CFT in which the tensor network is defined on a triangulation of hyperbolic space representing a fixed time slice of the AdS spacetime. Then, projecting the bulk legs onto a bulk state $\ket{\psi^{\text{bulk}}}$, one obtains a resulting boundary state which is a pure state in the Hilbert space $\mathcal{H}_\partial$:
\begin{equation}
\label{eq:TN_state}
    \ket{\psi} = \frac{1}{N} \left( \bra{\psi^{\text{bulk}}}\bigotimes_{\{xy\} \in \mathcal{E} \setminus \partial} \bra{xy} \right) \left( \bigotimes_{x \in \mathcal{V}} \ket{T_x} \right) \in \mathcal{H}_\partial ~,
\end{equation}
where the normalization constant $N$ ensures that $\braket{\psi | \psi} = 1$.

In an RTN, each tensor $\ket{T_x}$ is drawn at random using the Haar measure. This can be implemented by acting on a fixed reference state $\ket{0_x}$ with a Haar random unitary $U_x$. The advantage of this choice is that the statistical properties of Haar random unitaries let us compute various averaged quantities such as the entropy. In particular, by applying Schur’s lemma~\cite{harrow2013churchsymmetricsubspace}, one has:
\begin{equation}\label{eq:haar}
	\overline{(\ket{T_x}\bra{T_x})^{\otimes n}} = \int [dU_x] \left( U_x \ket{0_x} \bra{0_x} U_x^\dagger \right)^{\otimes n} \propto \sum_{g_x \in \mathcal{S}_n} g_x ~,
\end{equation}
where $g_x$ is an element of the symmetric group $\mathcal{S}_n$ which acts by permuting the ket edges of the $n$ copies of the state. 

This leads us to the computation of the entropy of a boundary subregion. Let the boundary Hilbert space be factorized as $\mathcal{H}_{\partial} = \mathcal{H}_A \otimes \mathcal{H}_{\bar{A}}$, where $A$ is the subregion of interest. The $n$-th R\'enyi entropy can be computed using a replica trick as the expectation value of a twist operator, which cyclically permutes the $n$ copies of $\mathcal{H}_A$ according to the permutation $\tau_n \in \mathcal{S}_n$. Using \Eqref{eq:haar}, the calculation in the RTN boils down to the partition function of a ferromagnetic Ising-like model with a sum over permutations at each vertex dictating the gluing pattern of the various replicas and the energy cost controlled by the Cayley distance between neighbouring permutations. The boundary conditions for this partition function are set by the replica trick, i.e., $g_x = \tau_n$ for $x \in A \subseteq \partial$, and $g_x = e$ (the identity element) for $x \in \bar{A} = \partial \setminus A$.

For the purposes of modeling AdS/CFT, we are interested in the large $\chi$ limit, which corresponds to a semiclassical $G\to0$ limit. In this limit, the entropy is self-averaging and thus can be approximately computed via the above replica trick. In this limit, the dominant contribution to the R\'enyi entropy arises from the ground state of the Ising-like model: a replica-symmetric configuration consisting of domains labeled by group elements $\tau_n$ and $e$, separated by a domain wall. In particular, the location of the domain wall is fixed by energy minimization. Thus, in this limit, we find:
\begin{equation}
\label{eq:bdry_IsingAction} 
\overline{{\rm Tr}\(\rho_{A}^n\)}    = e^{ -  (n-1) |\gamma_A| \ln \chi  }\,{\rm Tr}\left[(\rho^{{\text{bulk}}}_{E(A)})^n\right] ~,
\end{equation}
where $|\gamma_A|$ is the number of legs cut by the domain wall, and $\rho^{{\text{bulk}}}_{E(A)}$ is the partial density matrix associated with the bulk state $\ket{\psi^{\text{bulk}}}$ restricted to the $\tau_n$ domain in the ground state, labelled $E(A)$ since it will turn out to be the entanglement wedge of $A$. This yields:
\begin{equation}
\label{eq:bdry_renyi_entropy}
    S_n(A) \equiv \frac{1}{1-n} \log \[{\rm Tr}\(\rho_{A}^n\)\] = |\gamma_A| \ln \chi + S_n(\rho^{{\text{bulk}}}_{E(A)}) ~,
\end{equation}
where due to the self-averaging nature of this calculation, the above statement is true for almost all instances of the random tensors. The term proportional to $\ln \chi$ in \Eqref{eq:bdry_renyi_entropy} plays the role of the area term, while the second term corresponds to the matter entropy in the entanglement wedge. In the $n\to 1$ limit in particular, we obtain the RTN version of the Ryu-Takayanagi (RT) formula:
\begin{equation}
    S(A) = |\gamma_A| \ln \chi + S(\rho^{{\text{bulk}}}_{E(A)})=\S\(E(a)\),
\end{equation}
where we have defined an RTN version of the generalized entropy $\S$ associated to the generalized entanglement wedge.

\subsection{Generalized Entanglement Wedges}
\label{sub:genEW}
Having reviewed the computation of the boundary entropy in RTNs, we now propose a framework to define entropies associated to bulk regions. This defines the generalized entanglement wedge in RTNs and we show that it agrees with the BP proposal.

\begin{figure}[h]
    \centering
    \includegraphics[width=14cm]{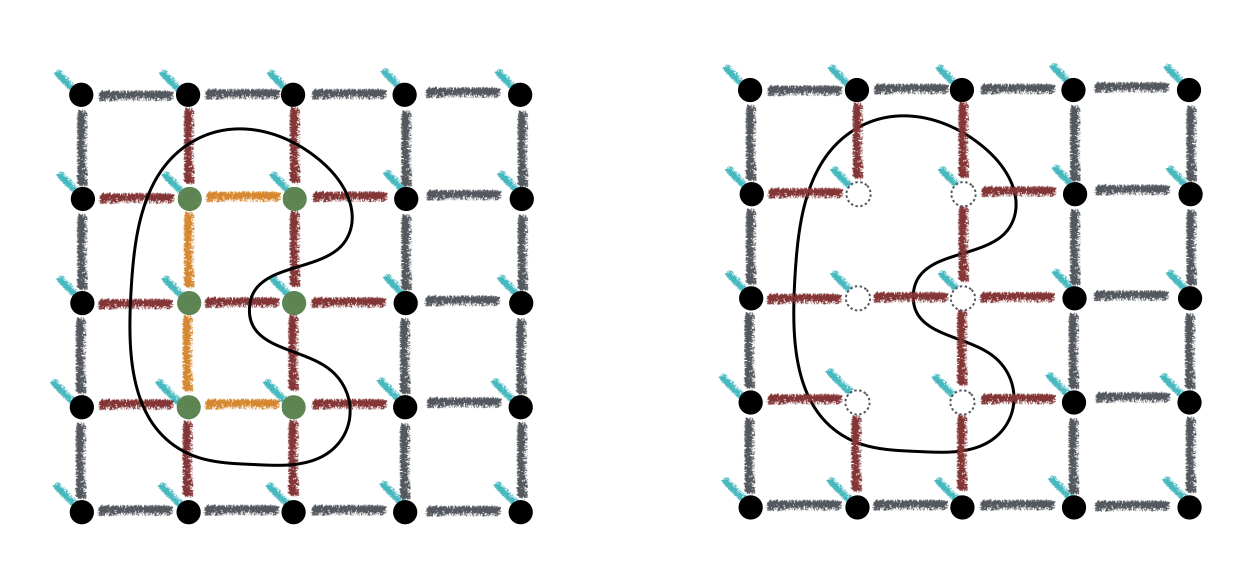}
    \caption{\label{fig:bulk-input-tn} A bulk input region in a tensor network. The green dots represent the vertices comprising the input region $a$. The yellow edges correspond to the internal edges $\E_a$, while the red edges represent the boundary $\partial a$ and the turquoise legs indicate the bulk legs. The right-hand panel shows the diagrammatic representation of the hollow-graphic state used to compute the entropy associated with $a$.}
\end{figure}

\begin{defn}
    \textbf{(Bulk Input Regions)} Given an RTN, an input region $a$ is defined as a subset of vertices, $a \subset \V$. The complement of $a$ is denoted by $\bar{a} = \V \setminus a$. \\
    The internal edges of $a$, denoted $\E_a$, consist of all edges that connect vertices within $a$ to each other. The boundary of $a$, denoted by $\partial a$, includes all in-plane edges connecting vertices in $a$ to $\bar{a}$. We denote the number of such boundary edges by $|\partial a|$. See \figref{fig:bulk-input-tn}.
\end{defn}

\begin{defn}
    \textbf{(Hollow-graphic State)} Given an RTN and an input region $a$, the generalized entanglement entropy associated with $a$ is defined using the hollow-graphic state:
    \begin{equation} \label{eq:TN_state_a}
    	\ket{\psi_a}=  \frac{1}{N_{a}} \left( \bra{\psi^{\text{bulk}}}\bigotimes_{\{xy\}\in 
        \E \setminus \partial \setminus \partial a \setminus \E_a
        }\bra{xy} \right) \left( \bigotimes_{x\in \V \setminus a} \ket{T_x} \right)\(\bigotimes_{x\in a}\ket{\phi^{\text{bulk}}_{x}} \)  ~,
    \end{equation}
    where $N_a$ is a normalization constant and $\ket{\phi^{\text{bulk}}_{x}}$ is a maximally entangled state on the bulk dangling leg at vertex $x$ and an auxiliary copy of it.
\end{defn}

Let us now explain how the hollow-graphic state is algorithmically constructed from the original RTN projected onto a bulk state. By removing tensors in $a$, as well as all its internal edges, we generate a new hollowed graph with open in-plane legs on $\pa a$ as well as open bulk legs in the region $a$. We add in an auxiliary copy for each bulk leg opened this way and put in a maximally entangled state $\ket{\phi^{\text{bulk}}_{x}}$ for each $x\in a$. This is a trick that allows us to flip bras coming from postselection on the bulk dangling legs in $a$ to kets.\footnote{There is an ambiguity in the precise choice of this state but it doesn't affect the entropies we are interested in computing.} The hollow-graphic state $\ket{\psi_{a}}$ thus obtained using \Eqref{eq:TN_state_a} is now defined on a larger Hilbert space $\mathcal{H}_{\pa}\otimes \mathcal{H}_{\pa a}\otimes \mathcal{H}_{a,\text{bulk}}$, where $\mathcal{H}_{\pa a}$ and $\mathcal{H}_{a,\text{bulk}}$ are the Hilbert spaces corresponding to the open in-plane legs on the boundary of $a$ and open bulk dangling legs in $a$ respectively. See \figref{fig:bulk-input-tn}. A similar construction has been proposed by Ref.~\cite{Akers:2025ahe}.\footnote{We thank Chris Akers for discussions on this point.} Note that this is a different construction from that provided in Ref.~\cite{Dong:2020uxp}, where one detaches the bulk legs but does not remove the tensors or in-plane legs resulting in a proposal for the effective entropy of bulk quantum fields.

Since in an RTN, there is no qualitative difference between open legs deep in the bulk and those at the boundary, we can use the previous calculation of the entropy as is. As in the case of boundary subregions, in the limit of large bond dimension, the dominant configuration corresponds to a domain wall homologous to the newly introduced boundary $\partial a$, resulting from the removal of $a$. The R\'enyi entropy of this state can be computed analogously with the boundary conditions imposed by setting $g_x = \tau_n$ for 
$x \in \partial a$, and $g_x = e$ for $x \in  \partial \setminus \partial a$ :
\begin{equation}\label{eq:bulk_IsingAction} 
	\overline{{\rm Tr}\(\rho_{a}^n\) } =  e^{ -  (n-1) |\gamma_a| \ln \chi  }\,{\rm Tr}\left[(\rho^{\text{bulk}}_{E(a)})^n\right]  ~,
\end{equation}
where $|\gamma_a|$ is the number of legs crossed by the energy minimizing domain wall, and $\rho^{\text{bulk}}_{E(a)}$ is the partial density matrix associated with the original bulk state $\ket{\psi^{\text{bulk}}}$ on the cyclically glued domain $E(a)$ of the tensor network, which will get identified with the generalized entanglement wedge of $a$. The resulting R\'enyi entropy is then
\begin{equation}
S_n(a) \equiv \frac{1}{1-n} \log\[{\rm Tr}\(\rho_{a}^n\) \] = |\gamma_a| \ln (\chi) + S_n(\rho^{\text{bulk}}_{E(a)}) ~.
\end{equation}
Again in the $n\to 1$ limit, we obtain
\begin{equation}
    S(a) = |\gamma_a| \ln \chi + S(\rho^{{\text{bulk}}}_{E(a)})=\S\(E(a)\).
\end{equation}
This is the analog of the RT formula for bulk subregions in RTNs. 

As anticipated above, the region $E(a)$ is precisely the \textit{generalized entanglement wedge}, providing an RTN realization of the BP proposal. In cases where the boundary of $a$ intersects the outer boundary of the tensor network, $\partial a \cap \partial \neq \varnothing$, the imposed boundary conditions ensure that $\partial E(a) \cap \partial = \partial a \cap \partial$.

\begin{rem}
  The case where the input region is a boundary subregion, i.e., $\partial a = A$ and $\E_{a} = \varnothing$, is a special instance of this prescription.
\end{rem}

Our analysis above also guarantees that all standard results for RTNs follow for the hollow-graphic state as well. There is not only an RT formula, but also a version of subregion duality, which involves acting on the open legs in $a$ in the hollow-graphic state to affect the semiclassical bulk degrees of freedom in $E(a)$.

Finally, let us comment on the map taking the original RTN state to the hollow-graphic state. Let $V_a$ denote the map corresponding to the removal of subregion $a$ from the tensor network, mapping the state $\ket{\psi}$ defined in \Eqref{eq:TN_state} to the state $\ket{\psi_a}$ defined in \Eqref{eq:TN_state_a}. Since the state $\ket{\psi}$ resides in a smaller Hilbert space compared to that of $\ket{ \psi_a}$, the map $V_a$ is an embedding. The inverse of this map is a projection from $\mathcal{H}_{\partial} \otimes \mathcal{H}_{\partial a}\otimes \mathcal{H}_{a,\text{bulk}}$ back to $\mathcal{H}_{\partial}$. This corresponds to ``sewing back in" the subregion $a$ into the tensor network by projecting onto an appropriate state in $\mathcal{H}_{\partial a}\otimes \mathcal{H}_{a,\text{bulk}}$. We denote this projection by $\Pi_a$. This acts on  $\ket{\psi_a}$ as 
\begin{equation}
    \Pi_a\ket{\psi_a} =  \Pi_a (V_a \ket\psi ) = \ket\psi~.
\end{equation}
These maps will play an important role in our following discussion on the independence of bulk subregions.

\subsection{Independence Relations and Entropy Inequalities}
\label{sub:ind}

Having understood how to make sense of generalized entanglement wedges for bulk regions, we can now discuss how generalized entanglement wedges work for multiple bulk regions, a prerequisite to discuss entropy inequalities like strong subadditivity (SSA). In quantum mechanics, SSA is stated as follows: given a state $\rho$ on a Hilbert space that can be factorized as $\mathcal{H} = \mathcal{H}^A \otimes \mathcal{H}^B \otimes \mathcal{H}^C$, the following inequality holds:
\begin{equation} 
    S(\rho^{AB}) + S(\rho^{BC}) \geq S(\rho^{B}) + S(\rho^{ABC}), 
    \label{ineq:SSA}
\end{equation}
where the density matrices $\rho^A = \mathrm{Tr}_{BC} \rho$ are defined through the partial trace. However, since the hollow-graphic state used to define the generalized entanglement wedge is subregion-dependent, as can be seen from \Eqref{eq:TN_state_a}, more care is required to achieve such a factorization.

In particular, we propose to define SSA in holography as follows: start by considering the hollow-graphic state $\ket{\psi_{abc}}$ for regions $a,b,c$ that are spacelike separated. In this hollowed RTN, it is manifest that the Hilbert spaces associated with the open legs of regions $a$, $b$ and $c$ factorize. The statement of SSA expressed using the RT formula then takes the form:
\begin{equation}
    \S(E(ab)|_{c'}) + \S(E(bc)|_{a'}) \geq \S(E(b)|_{(ac)'}) + \S(E(abc)),
    \label{ineq:SSA-EW}
\end{equation}
where $E(p)|_{q'}$ denotes the \emph{restricted entanglement wedge} of the bulk region $p$, defined as the entanglement wedge of $p$ computed in the spacetime $q'$, where the boundary of $q$ is treated analogously to the asymptotic boundary and the generalized entanglement wedge is not permitted to include any portion of $q$. Restricted entanglement wedges were first introduced in Ref.~\cite{Bousso:2025fgg} in the context of the complementarity of bulk holograms, and they arise naturally in our framework as well.

Notably \Eqref{ineq:SSA-EW} is different from the unrestricted version of SSA considered in previous work:
\begin{equation}
    \S(E(ab)) + \S(E(bc)) \geq \S(E(b)) + \S(E(abc)),
    \label{ineq:SSA-EW2}
\end{equation}
where one follows the standard BP proposal and doesn't impose any additional homology constraints. Nevertheless, if the following independence condition is satisfied for a collection of input regions $a_i$:\footnote{This independence condition is identical to the one used in the proof of the no-cloning theorem in Ref.~\cite{Bousso:2023sya}. That argument, as well as the time-dependent proofs of strong subadditivity ~\cite{Bousso:2023sya} and the monogamy of mutual information in ~\cite{Bousso:2024ysg}, r
elies on the assumption that the minimal and maximal entanglement wedges coincide for arbitrary unions of input subregions, i.e., $\emin = \emax$. This assumption is crucial: when $\emin \neq \emax$, the bulk dual of the von Neumann entropy is not simply given by the RT formula \cite{Akers:2020pmf}.}
\begin{equation}\label{eq:indep}
a_i \subset\left[ E\left( \bigcup_{j \neq i} a_j \right)\right]'~, 
\end{equation}
then it is guaranteed that $E(p)|_{q'}=E(p)$ since imposing the homology constraint doesn't affect the generalized entanglement wedge of $p$. Given this condition, Eq.~\eqref{ineq:SSA-EW} in fact reduces to Eq.~\eqref{ineq:SSA-EW2}. 

Similarly, we note that all the holographic entropy cone inequalities can be proved with the weaker assumption that all the $a_i$ are spacelike separated as long as the holographic entropy cone inequalities are expressed in terms of the areas of restricted entanglement wedges, as in Eq.~\eqref{ineq:SSA-EW}. Our framework suggests that this formulation is more natural compared to the unrestricted version previously considered since the generalized entropies of restricted entanglement wedges correspond to subregion entropies of a single state, rather than entropies of subregions in different states that are related through post-selection. Moreover, the unrestricted holographic inequalities previously considered follow as a special case of our restricted inequalities when the appropriate independence conditions \Eqref{eq:indep} are satisfied.

Now, \Eqref{eq:indep} was indeed imposed as an independence condition for all the unrestricted higher holographic entropy cone inequalities\footnote{Note that these higher holographic entropy cone inequalities hold only at leading order in $1/G$, when one considers only the area contribution to the generalized entropy. One can prove these inequalities when including bulk matter contributions only if further entropy inequalities are imposed on the bulk matter itself \cite{Akers:2021lms}.} in Refs.~\cite{Bousso:2024ysg,Bousso:2025mdp}. However, a puzzling feature of the BP proposal is that, in order for bulk subregions to satisfy strong subadditivity (SSA) in the unrestricted form of \Eqref{ineq:SSA-EW2}, only a weaker notion of independence was necessary. In both time-reflection symmetric and time-dependent settings, it suffices for input regions to be spacelike separated to ensure the unrestricted form of SSA. This is surprising since we have argued that \Eqref{ineq:SSA-EW} is the natural statement of SSA, which does not reduce to \Eqref{ineq:SSA-EW2} under this weaker independence condition.

\begin{figure}[ht]
    \centering
    \includegraphics[width=0.8\linewidth]{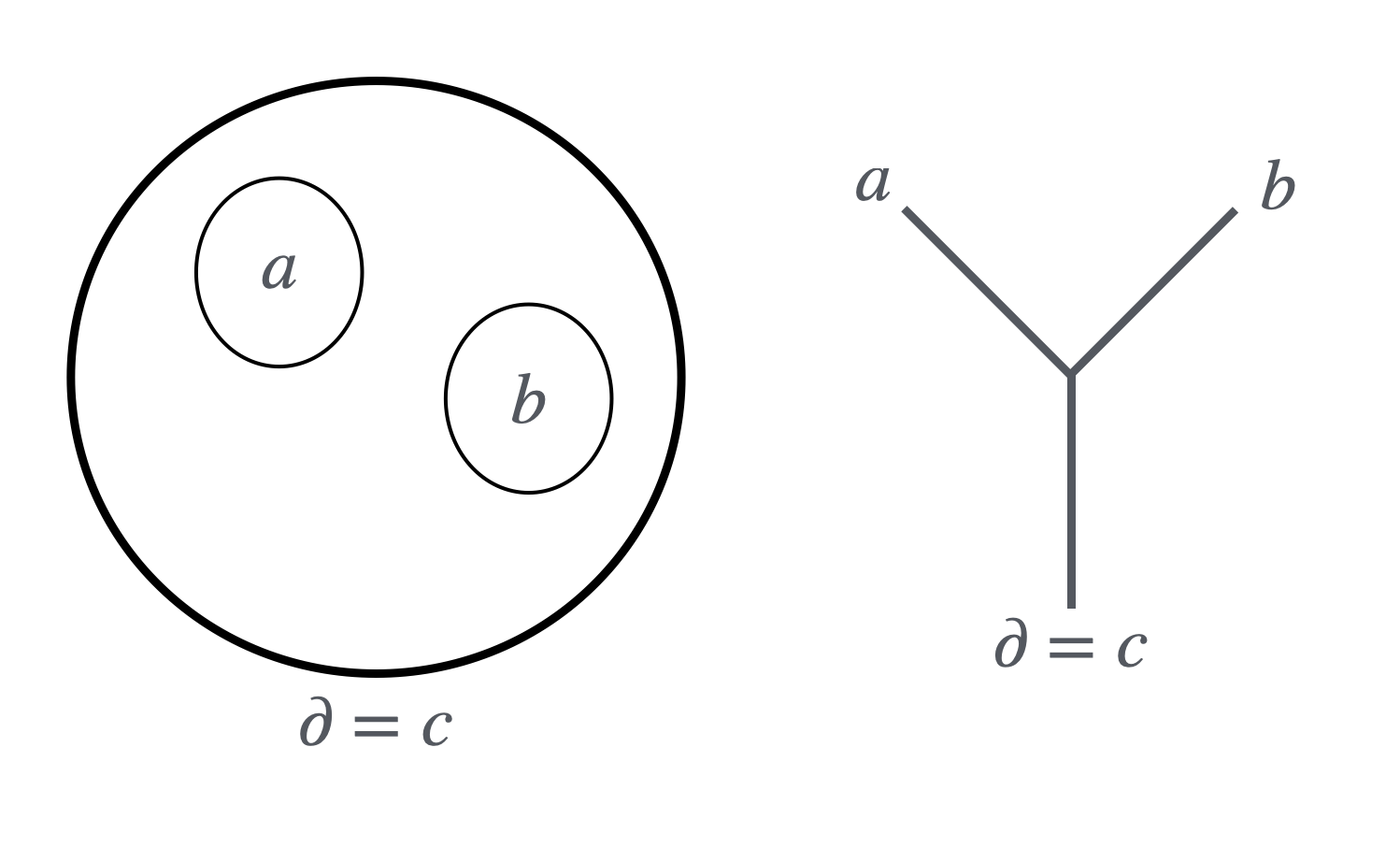}
    \caption{ \label{fig:GHZ} Schematic representation of the GHZ example. On the left, we see the tensor network analogue of the type of state we consider. The GHZ state is analogous to the state with subregions $a$ and $b$ both removed. The right side schematically represents the GHZ-type entanglement between respective subregions.}
\end{figure}

To resolve this confusion, our proposal here is that the fact that SSA holds under a weaker condition appears to be a special feature of semiclassical states with a good $G\to0$ limit. Given regions $a$, $b$, and $c$ in an RTN, the statement of \Eqref{ineq:SSA-EW2} in general situations amounts to a statement of SSA under postselection, which is not generally true in quantum mechanics. For an explicit example, consider the GHZ state (see Fig.~\ref{fig:GHZ}). In this case, we find $S(\rho^{ab}) = \ln(2)$. To compute $S(\rho^b)$ or $S(\rho^{bc})$, we first apply the postselection $\Pi_a$ which we take to be postselection onto $\ket{0}$. 
After this projection, we find $S(\rho^b) = 0$, and since the subsystem $bc$ is then in a pure state, we have $S(\rho^{bc}) = 0$. Similarly, $S(\rho^{abc}) = 0$. Putting this together, we see that the unrestricted version of SSA is violated as expected.

Now that we have seen why a stronger notion of independence is generally required, we still need to explain why, in the context of generalized entanglement wedges, a weaker notion of independence suffices specifically for strong subadditivity (SSA). This is a rather special and seemingly coincidental feature of SSA.

As noted above, in both the static case~\cite{Bousso:2022hlz} and the time-dependent case~\cite{Bousso:2023sya}, at leading order in $G$—where entropy is given by the area of the entanglement wedge in Planck units— no independence condition is required to prove SSA except that the input regions are spacelike. In the static setting, SSA follows from basic properties of Euclidean geometry, while in the time-dependent setting, the proof, though more intricate, remains purely geometric. The spacelike condition ensures that, in the intermediate step of the cut-and-paste construction, the resulting regions satisfy the appropriate homology constraints.\footnote{We thank Matthew Headrick for discussions that helped clarify this point.}

When including quantum corrections, care needs to be taken that the quantum state has a good $G\to0$ limit, or equivalently a good large bond dimension limit in the RTN. This is a restrictive set of states which appear to satisfy the unrestricted version of SSA, but our claim is that this is only a special feature of semiclassical holographic states and not a universal statement of quantum mechanics.

It is also important to note however that unrestricted SSA need not be an indicator of a holographic dual: similar behavior can be found in theories that are not known to possess a holographic dual. E.g., in many low-energy states of general theories where the entanglement entropy of a subregion follows an area law, SSA continues to hold even when the input regions significantly overlap.

\section{Holographic R\'enyi Entropy via Fixed-Geometry States}
\label{sec:holEE}

In this section, we review the calculation of R\'enyi entropy of a boundary subregion using the gravitational path integral. In particular, we will focus on a derivation where we decompose a holographic state into fixed-geometry states. In Refs.~\cite{Dong_2024,Penington:2024jmt}, this method was shown to agree with the computation by Ref.~\cite{Lewkowycz:2013nqa} for $n>1$ and rectify it for $n<1$. It will also serve as a natural formalism that allows us to generalize to the case of interest, i.e., the entropy of bulk regions. 

Consider a time-reflection symmetric holographic state $\ket{\psi}$ which is prepared by a standard Euclidean gravitational path integral with asymptotic boundary conditions.\footnote{For concreteness, we will consider setups with an asymptotic boundary relevant to AdS/CFT, but our analysis doesn't rely strongly on this as long as one can define the gravitational path integral in a sensible manner and choose appropriate gauge fixing conditions to specify the Cauchy slice on which we fix the geometry.} We denote the maximal volume slice anchored to the time-reflection symmetric slice on the boundary by $\Sigma$.  We will assume that the leading spacetime geometry corresponding to $\ket{\psi}$ preserves time-reflection symmetry and thus, $\Sigma$ is the time-reflection symmetric slice in this spacetime. This state can be decomposed into a basis of states $\ket{h}$ that have fixed geometry on the slice $\Sigma$ as\footnote{This is true in the approximation where we ignore wormhole effects, which in fact can become important when exponentially long times are involved. See, e.g., Ref.~\cite{Iliesiu:2024cnh} for the difference in JT gravity between the inner product in the disk approximation versus when wormhole effects are included.}
\begin{equation}\label{eq:state_sup_h}
    \ket{\psi} = \sum_{h} \psi(h)\ket{h}.
\end{equation}
The sum is over all possible gauge-inequivalent induced metrics $h$ on the time-reflection symmetric slice $\Sigma$ and $\psi(h)$ is a wavefunction that encodes the probability distribution of this superposition.


 \begin{figure}
    \centering
    \includegraphics[width=0.5\linewidth]{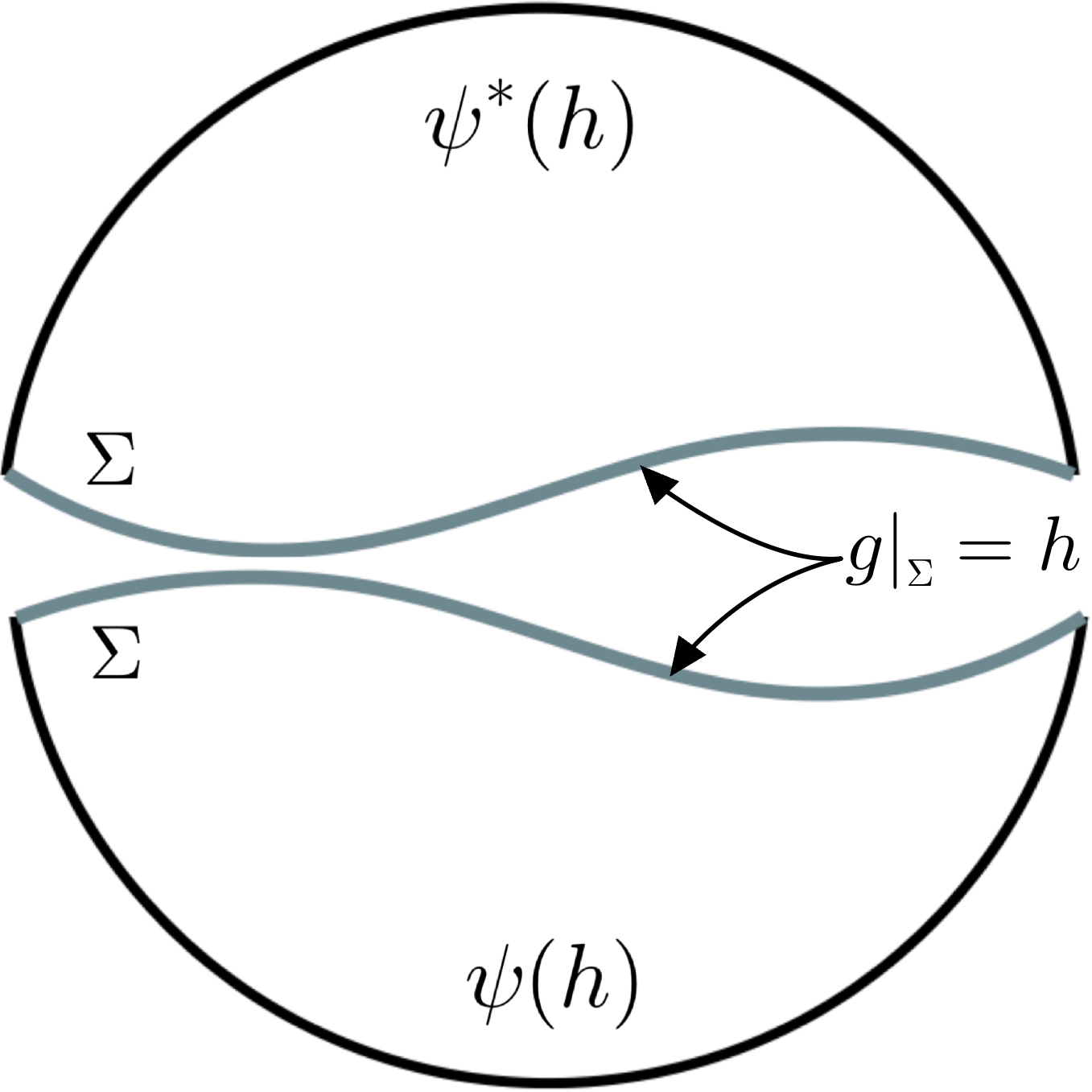}
    \caption{$\psi(h)$ is computed via a Euclidean gravitational path integral whose boundary conditions consist of the fixed geometry $h$ on the bulk slice $\Sigma$ and asymptotic AdS boundary conditions with sources. Gluing two copies along $\Sigma$ computes the norm $|\psi(h)|^2$ and the Euclidean geometry $g_h$ generically has a jump in extrinsic curvature across $\Sigma$. Integrating over $h$ computes the norm of the original state $\braket{\psi|\psi}=\int dh |\psi(h)|^2$, resulting in a smooth saddle $g_\psi$.}
    \label{fig:psi_h}
\end{figure}
 
The basis state $\ket{h}$ may be prepared by a Euclidean path integral with identical asymptotic boundary conditions to $\ket{\psi}$, but with an additional postselection that fixes the bulk spatial geometry on $\Sigma$ to be $h$. The postselection onto $h$ can be implemented by imposing a boundary condition on $\Sigma$ when computing the gravitational path integral, see \figref{fig:psi_h}. Since we have imposed a boundary condition on $\Sigma$, the Einstein equations need not be satisfied on $\Sigma$ and will generically involve jumps in extrinsic curvature.


The probability distribution $p(h)=|\psi(h)|^2$ can be calculated using the gravitational path integral and is given by
\begin{equation}\label{eq:path}
    |\psi(h)|^2 = \frac{\bra{\psi}\Pi_{h}\ket{\psi}}{\langle\psi|\psi\rangle} = \exp(I[g_\psi] - I[g_h]),
\end{equation}
where $\Pi_{h}$ is a projector which fixes the geometry on $\Sigma$ to be $h$ and $I$ is the bulk gravitational action. $g_\psi$ is the smooth Euclidean saddle computing the norm of the state $\ket{\psi}$, while $g_h$ is the Euclidean saddle computing the norm of the fixed-geometry state $\ket{h}$ which will generally have discontinuities in the extrinsic curvature at $\Sigma$. For semiclassical states, $\psi(h)$ is very sharply peaked around $h_\psi$, the induced metric on $\Sigma$ in the Euclidean saddle $g_\psi$. We give an explicit example of this decomposition for JT gravity in \secref{sec:eg}.

The reduced density matrix on a boundary subregion $A$ in this state is
\begin{equation}\label{eq:bdry_off_diag_dm}
    \rho_A= \sum_{h,h'} \psi(h)\psi^*(h')\mathrm{Tr}_{\partial\Sigma / A}\(\ket{h}\bra{h'}\).
\end{equation}
Thus, ${\rm Tr}(\rho_A^n)$ will receive both diagonal contributions where all $h=h'$ and off-diagonal contributions where some $h \neq h'$. For a semiclassical state with a sharply peaked wavefunction around a saddle point geometry, it was argued in Ref.~\cite{Dong:2023bfy} and shown in Ref.~\cite{Penington:2024jmt} that one can ignore the off-diagonal terms for the purpose of computing the R\'enyi entropy. Since only the diagonal terms lead to replica-symmetric saddles in the path integral, this also agrees with the standard assumption of replica symmetry.\footnote{This assumption of replica symmetry can be relaxed if, by assumption, there is a unique extremal surface that is homologous to the boundary subregion. See Ref.~\cite{Held:2024qcl} for more details.} Thus, using the diagonal approximation we have
\begin{equation}\label{eq:ZnA_general}
     {\rm Tr}\(\rho_A^n \) = \sum_h |\psi(h)|^{2n} \mathrm{Tr}\(\rho_{A}(h)^n\),
 \end{equation}
where $\rho_A(h)=\mathrm{Tr}_{\partial\Sigma / A}\(\ket{h}\bra{h}\)$ is the reduced density matrix on subregion $A$ in the fixed-geometry state $\ket{h}$.

The advantage of this method is that the computation of R\'enyi entropy in $\ket{h}$ is much easier than the problem of computing the R\'enyi entropy in $\ket{\psi}$.\footnote{While this is true for integer $n$, taking the $n\to1$ limit for fixed-geometry states is subtle since it does not commute with the $G\to0$ limit \cite{Dong:2023xxe}. Our interest here is to use this as an intermediate step for more general semiclassical holographic states, and in that case, we expect the limits to commute as is usually assumed in the literature.} To compute $\mathrm{Tr}\(\rho_{A}(h)^n\)$, we perform the replica trick. The boundary conditions involve $n$ copies of the asymptotic boundary glued together in the standard fashion with subregion $A$ glued cyclically across the $n$ copies, as well as a fixed-geometry boundary condition in the bulk on each of the copies. As usual, the states prepared by the Euclidean path integral are not normalized and thus, we must fix the normalization as:
\begin{equation} \label{eq:bdry_znormalization}
    \mathrm{Tr}\(\rho_{A}(h)^n\) = \frac{Z_n(A,h)}{Z_1(h)^n}~,
\end{equation}
where $Z_n(A,h)$ is the replica partition function for the fixed-geometry state $\ket{h}$ and $Z_1(h)=\braket{h|h}$.

In the semiclassical limit, the replica partition function can be evaluated using a saddle point approximation. For general states, this is challenging due to the difficulty in finding the corresponding saddles. However, the advantage of using fixed-geometry states is that the saddles take a simpler form. Since the bulk geometry on $\Sigma$ in each copy is already measured, all one needs to do is take $n$ copies of the bulk geometry computing $Z_1$ and glue together the fixed-geometry slices of the $n$ bras and $n$ kets. 

\begin{figure}
    \centering
    \includegraphics[width=1.0\linewidth]{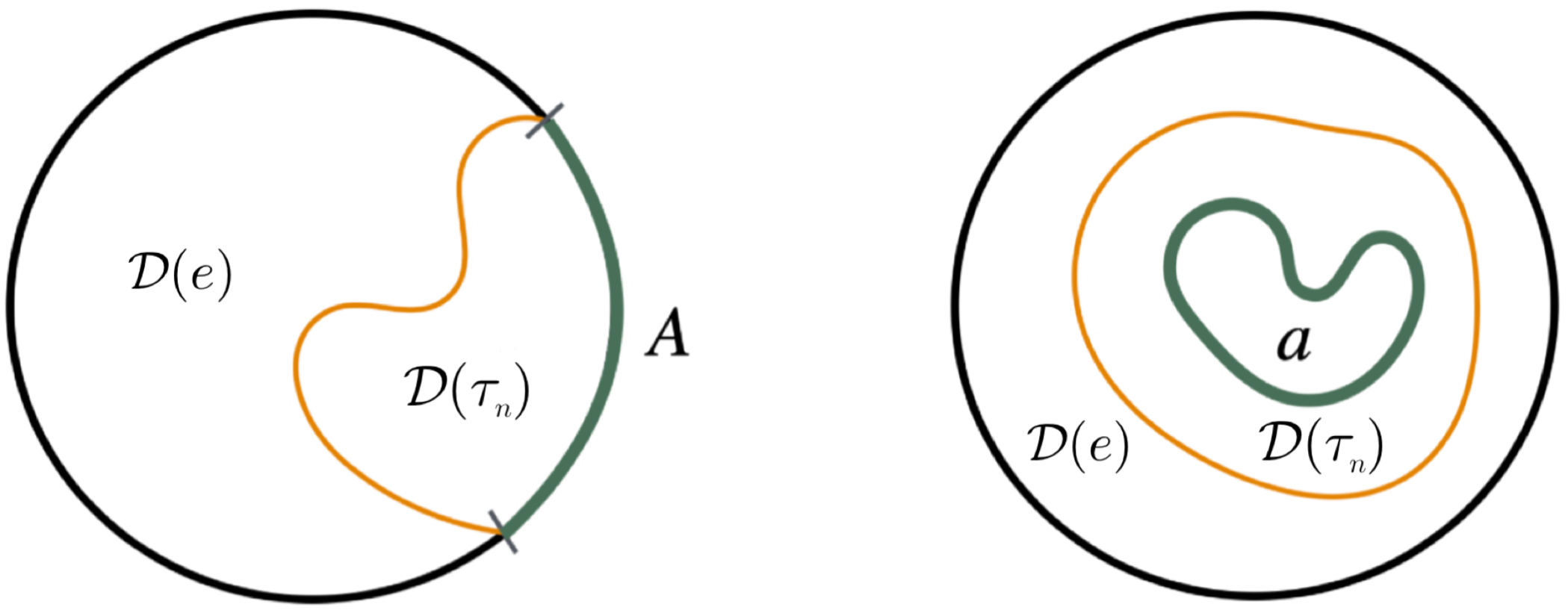}
    \caption{A figure representing the domain wall configurations contributing to the action when the input is a boundary subregion (left) and a bulk subregion (right).}
    \label{fig:domainwalls}
\end{figure}

The choice of gluing corresponds to selecting a map from the bulk spatial geometry to the permutation group $\mathcal{S}_n$. This map divides the bulk geometry into domains $\mathcal{D}(g)$ for $g \in \mathcal{S}_n$, with the $n$ copies of bulk regions being glued according to permutation $g$ in the region $\mathcal{D}(g)$, subject to the appropriate boundary conditions. Moreover, the asymptotic boundary conditions are fixed by the replica trick. The boundary subregion $A$ is assigned the cyclic permutation $\tau_n$ in the boundary replica trick and thus, the bulk region $\mathcal{D}(\tau_n)$ will have $A$ as its conformal boundary. Similarly, the boundary subregion $\bar{A}$ is assigned the identity permutation $e$. It should now be clear that this is identical to the configurations showing up in the RTN calculation, see \figref{fig:domainwalls}.

To find the saddle points, we must identify the configuration that minimizes the total action, which receives contributions from both the matter fields and the gravitational sector. The gravitational action itself has two components. The first arises from the domain walls separating the different domains $\mathcal{D}(g)$ on the fixed-geometry slice, while the second comes from everywhere else. 

In regions away from the domain walls, the contribution to the action is given by $n$ times the action of the saddle point that computes $Z_1(h) = \braket{h|h}$. This part of the action cancels against the normalization factor appearing in \Eqref{eq:bdry_znormalization}.

\begin{figure}
    \centering
    \includegraphics[width=1\linewidth]{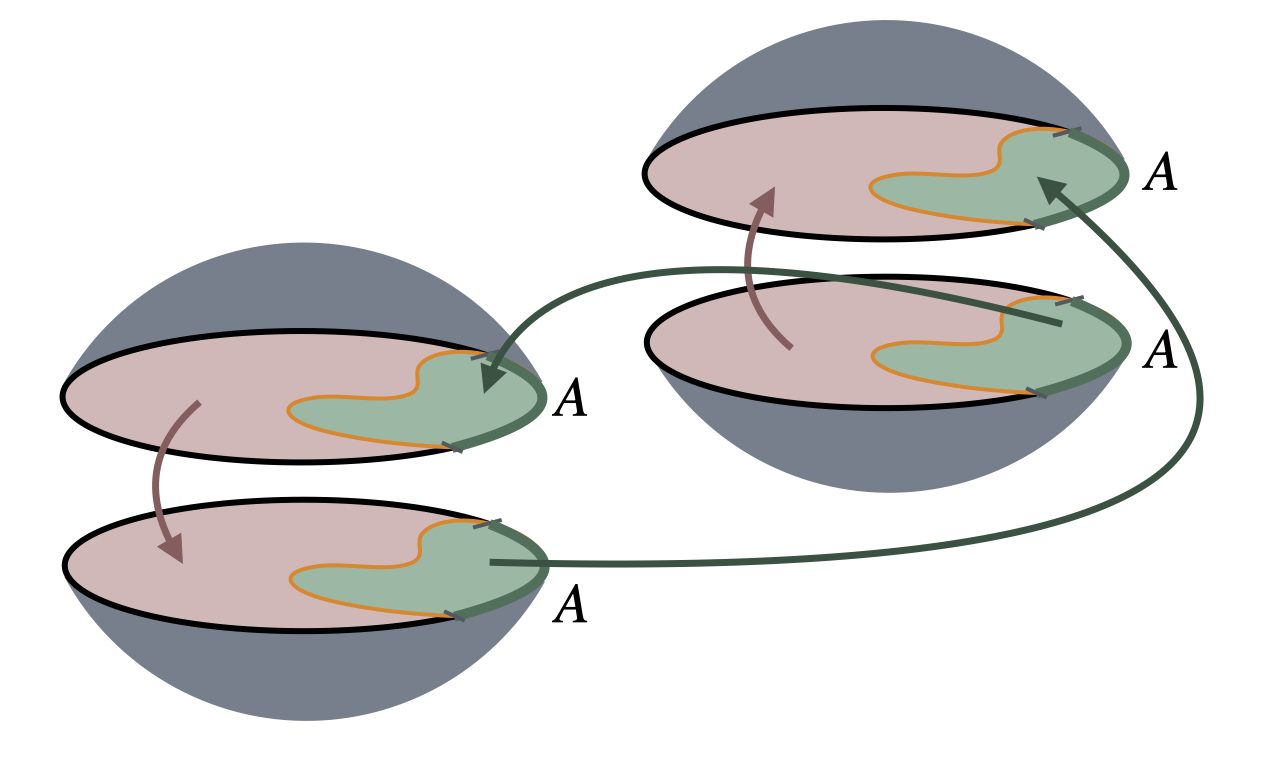}
    \caption{The dominant contribution to the replica path integral comes from geometries in which the regions of the bulk slice $\Sigma$ within the domain wall (sharing a conformal boundary with A) are glued cyclically, while the regions outside the domain wall are glued identically.  }
    \label{fig:Z2_bdry_gluing}
\end{figure}

The remaining contribution to the gravitational action arises from the conical singularities on the domain walls. In a fashion similar to the RTN calculation, the only domains that appear in the minimal action configuration are $e$ and $\tau_n$, consistent with the standard assumption of replica symmetry in holography \cite{Lewkowycz:2013nqa}. See \figref{fig:Z2_bdry_gluing} for the corresponding Euclidean geometry. The action associated with this conical singularity is given by
\begin{align}
I_{\text{con}} = ({ n - 1})\frac{\mathcal{A}(\partial \mathcal{D}(\tau_n),h)}{4 G_N} 
\end{align}
where $\mathcal{A}(\partial \mathcal{D}(\tau_n),h)$ is the area of the domain wall between $\mathcal{D}(e)$ and $\mathcal{D}(\tau_n)$ in geometry $h$. For more details on this derivation, see Refs.~\cite{Dong:2018seb, Akers:2018fow, Penington:2019kki}. The matter contribution just comes from computing the $n$th moment of the reduced bulk density matrix on $\mathcal{D}(\tau_n)$, similar to the RTN calculation.

Combining the two contributions and minimizing the resulting action, we have
\begin{equation}\label{eq:tr_rhon_h}
   \mathrm{Tr}\(\rho_{A}(h)^n\) = e^{-(n-1) \frac{\mathcal{A}(\gamma_A,h)}{4 G}} {\rm Tr}\left[\left(\rho^{\text{bulk}}_{E(A)}(h)\right)^n\right] ,
\end{equation}
where $\gamma_A$ is the RT surface of $A$ in geometry $h$, $E(A)$ is the corresponding entanglement wedge and $\rho^{\text{bulk}}(h)$ is the bulk density matrix of the matter fields in the fixed-geometry state. We have explicitly expressed the area of the RT surface as a function of $h$ to clarify its dependence on the induced metric on $\Sigma$. 

Using \Eqref{eq:tr_rhon_h} in \Eqref{eq:ZnA_general}, we obtain:
\begin{equation}\label{eq:ZmA}
    {\rm Tr}( \rho_A^n ) = \sum_h |\psi(h)|^{2n} e^{-(n-1) \frac{\mathcal{A}(\gamma_A,h)}{4 G}}{\rm Tr}\left[\left(\rho^{\text{bulk}}_{E(A)}(h)\right)^n\right].
\end{equation}
Finally, we can perform a saddle point approximation over $h$ to obtain
\begin{equation}
    {\rm Tr}( \rho_A^n ) \approx |\psi(h_n)|^{2n} e^{-(n-1) \frac{\mathcal{A}(\gamma_A,h_n)}{4 G}}{\rm Tr}\left[\left(\rho^{\text{bulk}}_{E(A)}(h_n)\right)^n\right],
\end{equation}
where $h_n$ is the saddle point geometry of \Eqref{eq:ZmA}. Note that $h_1 = h_\psi$ and the shift in saddle as a function of $n$ reproduces the expected backreaction for R\'enyi entropies \cite{Dong:2023bfy}. The R\'enyi entropy is given by
\begin{equation} \label{eq:renyi_A}
    S_n(A)   = \frac{2n}{n-1} \log\left(\frac{|\psi(h_{\psi})|}{|\psi(h_n)|}\right) + \frac{\mathcal{A}(\gamma_A,h_n)}{4G} + S_n\left(\rho^{\text{bulk}}_{E(A)}(h_n)\right).
\end{equation}
The von Neumann entropy is obtained by taking the $n \rightarrow 1$ limit of the preceding expression, yielding the RT formula
\begin{equation}
    S(A) = \lim_{n \rightarrow 1} S_n(A) = \frac{\mathcal{A}(\gamma_A,h_\psi)}{4G} + S\left(\rho^{\text{bulk}}_{E(A)}({h_\psi})\right),
\end{equation}
where we have made the dependence on the bulk metric explicit.


\section{BP Proposal from the Gravitational Path Integral}
\label{sec:main}

In order to define the entropy of bulk regions, we will now follow a strategy similar to the one employed above for boundary regions. The important new ingredient will be to propose a definition of the hollow-graphic state using the gravitational path integral in the basis of fixed-geometry states. Once we do that, the calculation of the R\'enyi entropy follows by employing a diagonal approximation. In the $n\to 1$ limit, we will show that this gives us the BP proposal.

Before we start, it is useful to discuss an important conceptual issue. To define a bulk subregion, we must specify it in a gauge-invariant manner. In particular, to define a bulk subregion in a general holographic state $\ket{\psi}$, which is a superposition of fixed-geometry states $\ket{h}$, we must provide a gauge-invariant definition of the subregion on all geometries $h$. \emph{A priori}, there are various ways to define such regions gauge-invariantly. Our analysis in the following will remain agnostic about the particular way in which the region is specified, especially because we will recover the BP proposal in the $n\to1$ limit independently of these details. For explicit examples of different ways to gauge-fix a subregion, see \secref{sec:eg}. 

\begin{figure}
    \centering
    \includegraphics[width=0.8\linewidth]{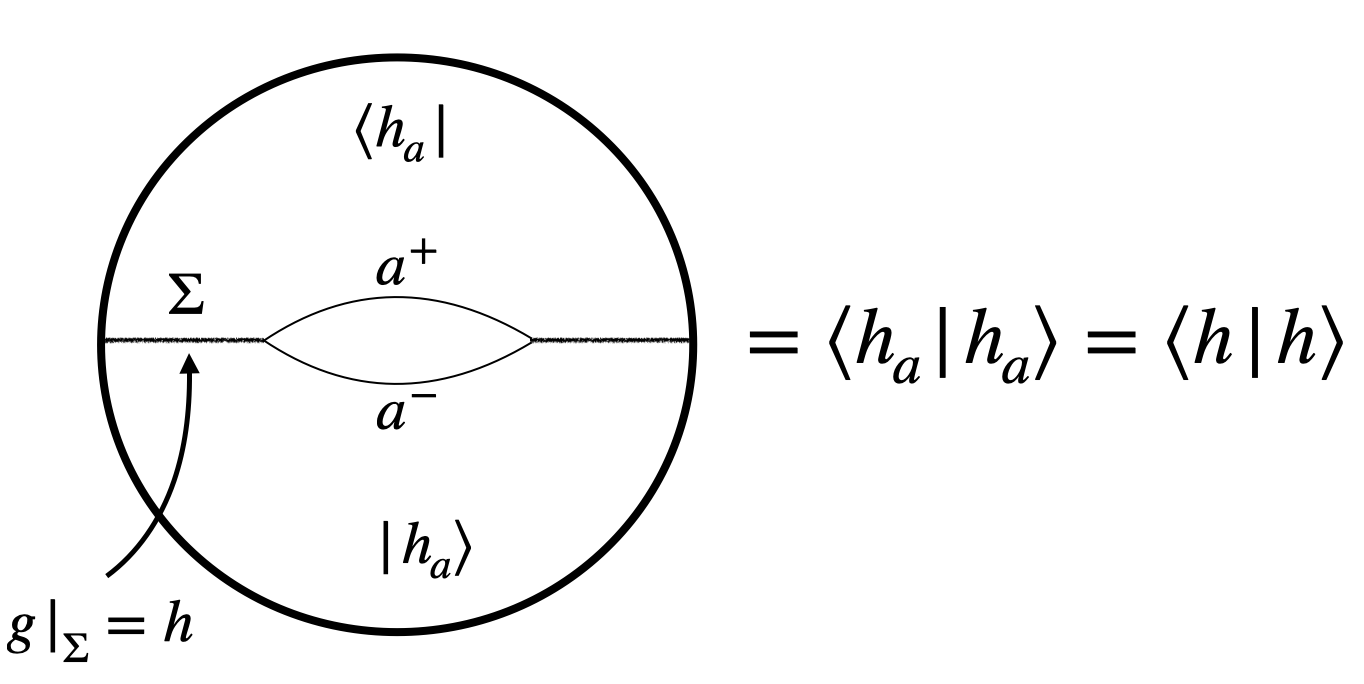}
    \caption{A schematic depiction of the Euclidean path integral with additional boundaries $a^-$ and $a^+$. When the induced metric on $a^-,a^+$ is chosen to agree with $h$ on region $a$, the Euclidean spacetime is the original fixed-geometry saddle with an infinitesimal-width cut around the subregion $a$. The hollow-graphic map is norm-preserving.}
    \label{fig:h_mag}
\end{figure}

Let $a$ be an input wedge specified in some gauge-invariant manner and $\ket{h}$ a fixed-geometry state. To define the entropy of this bulk region, we will take inspiration from the RTN and construct a hollow-graphic state. In the case of the RTN, the hollowing map $V_a$ was defined by removing the tensors in that region. In the case of gravitational states, we propose to hollow it out by removing a tiny slit around the region $a$ to introduce new boundaries $a^+$ and $a^-$, see \figref{fig:h_mag}. These new boundaries are additional locations where boundary conditions can be specified and thus describe an additional Hilbert space factor.\footnote{A similar construction was proposed in the context of cosmology by Ref.~\cite{Ivo:2024ill}.} 

In general, the induced metric on $a^-,a^+$ can be specified independent of the original $h$. When we choose the induced metric on $a^-,a^+$ to agree with the original $h$, we obtain the proposed hollow-graphic state $\ket{h_a}$. This geometric construction defines the map $V_a$:
\begin{equation}
    \ket{h_a} = V_a \ket{h}.
\end{equation}
From \figref{fig:h_mag}, it is also clear that the Euclidean action remains unchanged under removal of this infinitesimal slit and thus, the norm is preserved, i.e., $\braket{h_a|h_a}=\braket{h|h}$. 

We can then linearly extend the action of $V_a$ to a general holographic state by decomposing it in the basis of fixed-geometry states:
\begin{equation}
    \ket{\psi_a} =V_a \ket{\psi} = \sum_h  \psi(h)\,\ket{h_a} ~,
\end{equation}
where we have used the decomposition obtained in \Eqref{eq:state_sup_h}. We remind the reader that this map is well-defined and unique as long as the states $\ket{h}$ must form a basis of the Hilbert space. Although this is not strictly true due to wormhole corrections, we will work in a regime where they provide an approximate basis as we demonstrate in our examples in \secref{sub:JT}.


We can now obtain the reduced density matrix for subregion $a$ by tracing out its complement on $\partial \Sigma$:
\begin{equation}
    \rho_a = \mathrm{Tr}_{\partial  \Sigma \setminus \partial a} \left( V_a \rho V_a \right)=  \sum_{h,h'} \psi(h)\psi^*(h') 
    \mathrm{Tr}_{\partial  \Sigma \setminus \partial a}\left(  
    \ket{h_a} \bra{h_a'} \right) 
    ~.
\end{equation}
Using this, we can compute ${\rm Tr}( \rho_a^n )$ and just as in the boundary subregion case, there are both diagonal and off-diagonal terms that contribute to it. Motivated by replica symmetry, we will again assume a diagonal approximation and ignore the off-diagonal terms to obtain
\begin{equation}\label{Zna_general}
     {\rm Tr}( \rho_a^n ) = \sum_h |\psi(h)|^{2n}\, \mathrm{Tr}\(\rho_{a}(h)^n\),
\end{equation}
where $\rho_{a}(h)$ is the reduced density matrix on $a$ in the fixed-geometry hollow-graphic state $\ket{h_a}=V_a \ket{h}$, i.e.,
\begin{equation}
    \rho_a(h) 
    = \mathrm{Tr}_{\partial  \Sigma \setminus \partial a} \left[  \ket{h_a} \bra{h_a}  \right]~.
\end{equation}

\begin{figure}
    \centering
    \includegraphics[width=1\linewidth]{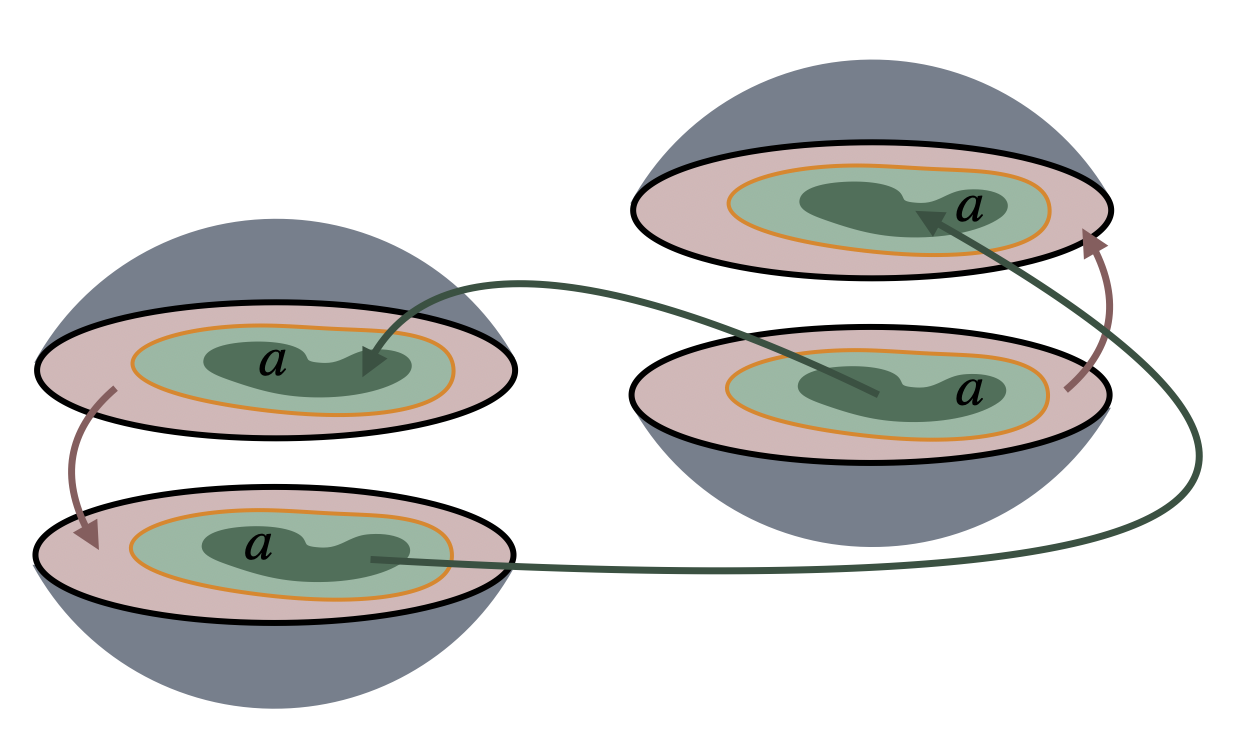}
    \caption{The dominant contribution to the replica path integral comes from geometries in which the regions of the bulk slice $\Sigma$ within the domain wall containing $a$ are glued cyclically, while the regions outside the domain wall are glued identically. }
    \label{fig:Z2_bulk_gluing}
\end{figure}

As in the boundary subregion case, $\mathrm{Tr}\(\rho_{a}(h)^n\)$ is computed using the gravitational path with gluing rules analogous to those in the tensor network calculation. Namely, we divide the bulk slice $\Sigma$ into domains $\mathcal{D}(g)$ associated with permutations $g\in \mathcal S_n$ and glue $n$ copies of $\Sigma$ with appropriate boundary conditions. As in the tensor network calculation, the boundary $\partial a$ is assigned the cyclic permutation $\tau_n$, and $\partial\Sigma / \partial a$ is assigned the identity permutation $e$. See \figref{fig:domainwalls} and \figref{fig:Z2_bulk_gluing}. 

As before, the saddle-point configuration preserves replica symmetry and consists only of two domains, $\mathcal{D}\(\tau_n\)$ and $\mathcal{D}\(e\)$. The action of the saddle-point configuration is given by the conical singularity along the domain wall $\pa \mathcal{D}(\tau_n)$. Including the contribution from the bulk density matrix, we find 
\begin{equation}\label{eq:Zma}
    \mathrm{Tr}(\rho_a^n ) = \sum_h |\psi(h)|^{2n} e^{-(n-1) \frac{\mathcal{A}(\gamma_a,h)}{4G}}{\rm Tr}[ (\rho^{\text{bulk}}_{E(a)}(h))^n]
 ~,
\end{equation}
where $\mathcal{A}(\gamma_a,h)$ is the area of the surface homologous to $\partial a$ in the geometry $h$ that minimizes the generalized entropy of the region $E(a)$ bounded by it.
 
The dominant saddle of \Eqref{eq:Zma} is denoted $h_n$, with the special case $h_1=h_\psi$, the leading geometry of the state $\ket{\psi}$. From the form of the action, it is clear that the precise location of $h_n$ depends on $n$. Moreover, it also depends on how the subregion $a$ is gauge-invariantly specified since this affects the action functional being optimized. Thus, we have
\begin{equation}
    \mathrm{Tr}(\rho_a^n ) \approx |\psi(h_n)|^{2n} e^{-(n-1) \frac{\mathcal{A}(\gamma_a,h_n)}{4G}} {\rm Tr}\left[\left (\rho^{\text{bulk}}_{E(a)}({h_n})\right)^n\right]
~,
\end{equation}
and the Rényi entropy becomes
\begin{equation}\label{eq:renyi_a}
    S_n(a) = \frac{2n}{n-1} \log\left( \frac{|\psi(h_\psi)|}{|\psi(h_n)|} \right) + \frac{\mathcal{A}(\gamma_a,h_n)}{4G}+ S_n \left (\rho^{\text{bulk}}_{E(a)}({h_n})\right)
 ~,
\end{equation}
where $h_n$ carries implicit $n$-dependence. 

Finally, taking the $n\to1$ limit, and recalling that $h_1=h_\psi$, we have
\begin{equation}\label{eq:BPproof}
    S(a) = \lim_{n \to 1} S_n(a) = \frac{\mathcal{A}(\gamma_a,h_\psi)}{4G} +S\left(\rho^{\text{bulk}}_{E(a)}({h_\psi})\right) ~.
\end{equation}
Importantly, the dependence on the precise way in which the bulk region is defined gauge-invariantly now drops out since all such prescriptions are chosen to agree on the leading geometry $h_\psi$. Thus, we have derived the BP proposal \Eqref{eq:BPproof} from the gravitational path integral. Moreover, we see that the entanglement wedge $E(a)$ in the geometry $h_\psi$ is independent of the specific gauge-invariant prescription used to extend the definition of $a$ to other geometries. This is consistent with the fact that Refs.~\cite{Bousso:2022hlz,Bousso:2023sya} didn't need to ever specify a particular gauge-invariant definition of bulk regions.

\section{Examples}
In this section, we will demonstrate how our formalism works in concrete examples where we compute the R\'enyi entropies for bulk regions on the time-symmetric slice in the thermofield double state. In particular, we will consider JT gravity, and higher-dimensional Einstein gravity in the minisuperspace approximation where we restrict to spherically symmetric configurations. A brief outline of the procedure is as follows:

\begin{enumerate}[i)]
    \item\label{step1} We begin with the semiclassical state $\ket\psi = \ket{TFD(\beta)}$ corresponding to a two-sided black hole of temperature $1/\beta$. The dominant geometry on the time-reflection symmetric slice $\Sigma$ is denoted $h_\psi$. We then consider a region $a$ on $\Sigma$.
    \item\label{step2} We expand $\ket{\psi}$ in an (approximate) basis of fixed-geometry states $\ket{h}$. In  JT gravity, we use fixed-dilaton states $\ket{\Phi_h}$, while in Einstein gravity, we work with states of fixed horizon area $\ket{\mathcal{A}_h}$.\footnote{In JT gravity, the entire geometry and dilaton profile is determined by the value of $\Phi$ at the horizon due to the constraint equations, see \Eqref{eq:JTSlice2}. Similarly, in the spherically symmetric minisuperspace approximation to Einstein gravity, fixing the horizon area fixes the entire geometry} In either case, the wavefunction $\psi(h)$ is given by the Hartle-Hawking wavefunction in the relevant basis.
    \item\label{step3} We then gauge-invariantly specify the region $a$ in every geometry $h$. For each $h$, we use the gravitational path integral to compute the R\'enyi entropy of $a$. 
    \item\label{step4} Finally, we compute the saddle-point in the integral over geometries in \Eqref{eq:Zma} to obtain the R\'enyi entropy $S_n(a)$ in the semiclassical state $\ket{\psi}$. As $n\rightarrow 1$, this reproduces the BP proposal: $S(a) = \frac{\mathcal{A}(\gamma_a,h_\psi)}{4G}$, where $\mathcal{A}(\gamma_a,h_\psi)$ is the minimal area surface homologous to $a$ in the eternal black hole geometry. 
\end{enumerate}

By providing various different methods of specifying the region $a$, we will show explicitly how, for $n>1$, the R\'enyi entropies depend on this choice, but as $n\rightarrow1$, the dependence vanishes giving an entanglement entropy consistent with the BP proposal.   

\label{sec:eg}
\subsection{JT Gravity}
\label{sub:JT}
To address (\ref{step1}) and (\ref{step2}) above, we now briefly present the Hartle-Hawking wavefunction $\psi(h)$ for the two-sided geometry. For details of the computation, largely following Ref.~\cite{Harlow:2018tqv}, see \ref{sec:App_HH}. Then we will address (\ref{step3}) and (\ref{step4}) by providing different gauge-invariant specifications for an interval in one of the exterior regions, and by computing the corresponding saddles for the R\'enyi entropy. 

\begin{figure}
    \centering
\includegraphics[width=0.5\linewidth]{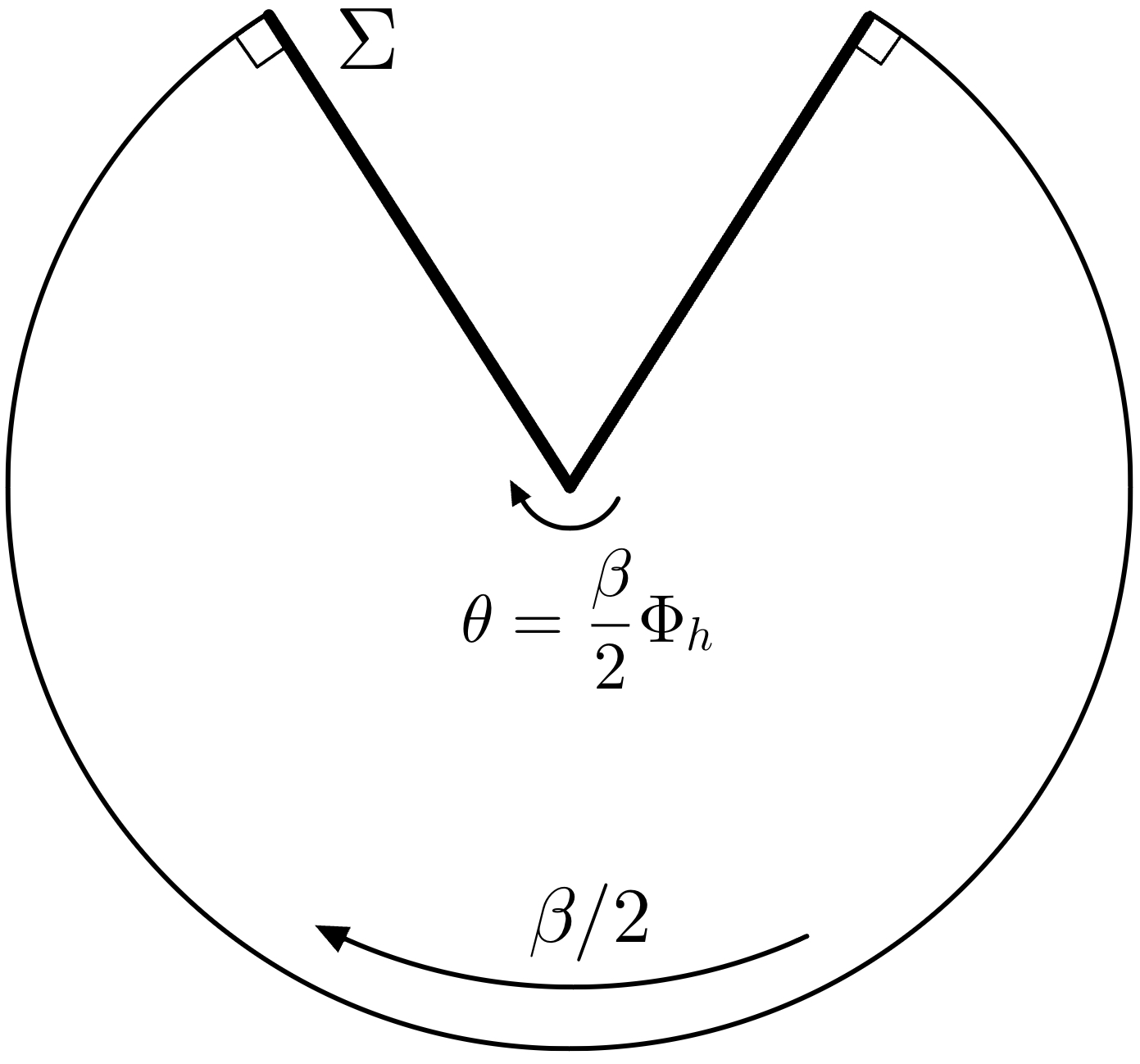}
    \caption{The Hartle-Hawking wavefunction is computed using a saddle-point geometry $g_{\Phi_h}$ with an asymptotic boundary of length $\frac{\beta}{2}$ and a boundary $\Sigma$ with the specified dilaton value $\Phi_h$ at the horizon. At generic values of $\Phi_h$, there is a kink in the slice $\Sigma$.}
    \label{fig:JT_psi_h}
\end{figure}

We begin with the thermofield double state prepared by a Euclidean evolution $\beta/2$, labelled $\ket{\psi_{\beta/2}}$. This Hartle-Hawking state can be expanded in the fixed-dilaton basis as $\ket{\psi_{\beta/2}}=\sum_{\Phi_h}\psi_{\beta/2}(\Phi_h)\ket{\Phi_h}$, where $\Phi_h$ denotes the value of the dilaton at the horizon. This provides a complete basis for the Hilbert space in the approximation where we ignore wormhole contributions. The wavefunction $\psi_{\beta/2}(\Phi_h)=\braket{\Phi_h|\psi_{\beta/2}}$ is constructed using the Euclidean gravitational path integral with an asymptotic AdS boundary $\partial M$ of length $\beta/2$, and an additional codimension-1 boundary in the bulk given by a slice $\Sigma$ with the specified dilaton value at the horizon. See figure \ref{fig:psi_h}. The saddle-point geometry $g_{\Phi_h}$ of this action is a section of the two-sided black hole geometry with a kink of angle $\theta = \frac{\beta}{2}\Phi_h$ at the horizon, depicted in \figref{fig:JT_psi_h}. Evaluating the action, the wavefunction is found to be
\begin{equation}\label{eq:JTHH}
\psi_{\beta/2}(\Phi_h)\approx e^{-I_E[g_{\Phi_h}]}= \exp{\{\pi(\Phi_0 + \Phi_h)-\frac{\beta}{4}
 \Phi_h^2\}}.
 \end{equation}

Having decomposed $\ket{\psi_{\beta/2}}$ into fixed-dilaton states $\ket{\Phi_h}$, we are now interested in computing the entropy of an interval on $\Sigma$. To discuss this, it is useful to write down the metric and dilaton profile on $\Sigma$ for each choice of $\Phi_h$:
\begin{equation}
    ds^2=\frac{dx^2}{1+x^2}\qquad \Phi=\Phi_h\sqrt{1+x^2},
\end{equation}
where the slice $\Sigma$ is parametrized by coordinate $x\in \(-\infty,\infty\)$ and the horizon is at $x=0$.

Let $A$ and $B$ denote the spacetime points at the boundaries of $a$, so that $\partial a = A\cup B$. Generically, for any gauge-invariant specification of the endpoints, there are two cases to consider: 
\begin{itemize}
    \item[I)] The endpoints $A$ and $B$ both lie on the same side of the horizon, for example $x_B>x_A>0$.
    \item[II)] The endpoints lie on opposite sides of the horizon, so that $a$ contains the bifurcation point. For example, we may have $x_A<0< x_B$.
\end{itemize}
One also needs to be able to specify which side of the horizon the points $A$ and $B$ lie on because of the exact symmetry $\mathbb{Z}_2$ in this problem. This can be done by using the asymptotic boundaries as a reference.\footnote{To illustrate this point, we note that specifying the value of $\Phi$ determines $|x|$, but not its sign.  The sign can be additionally determined by specifying whether or not the endpoint is on the same side as a chosen boundary.}
\begin{figure}
    \centering
    \includegraphics[width=0.9\linewidth]{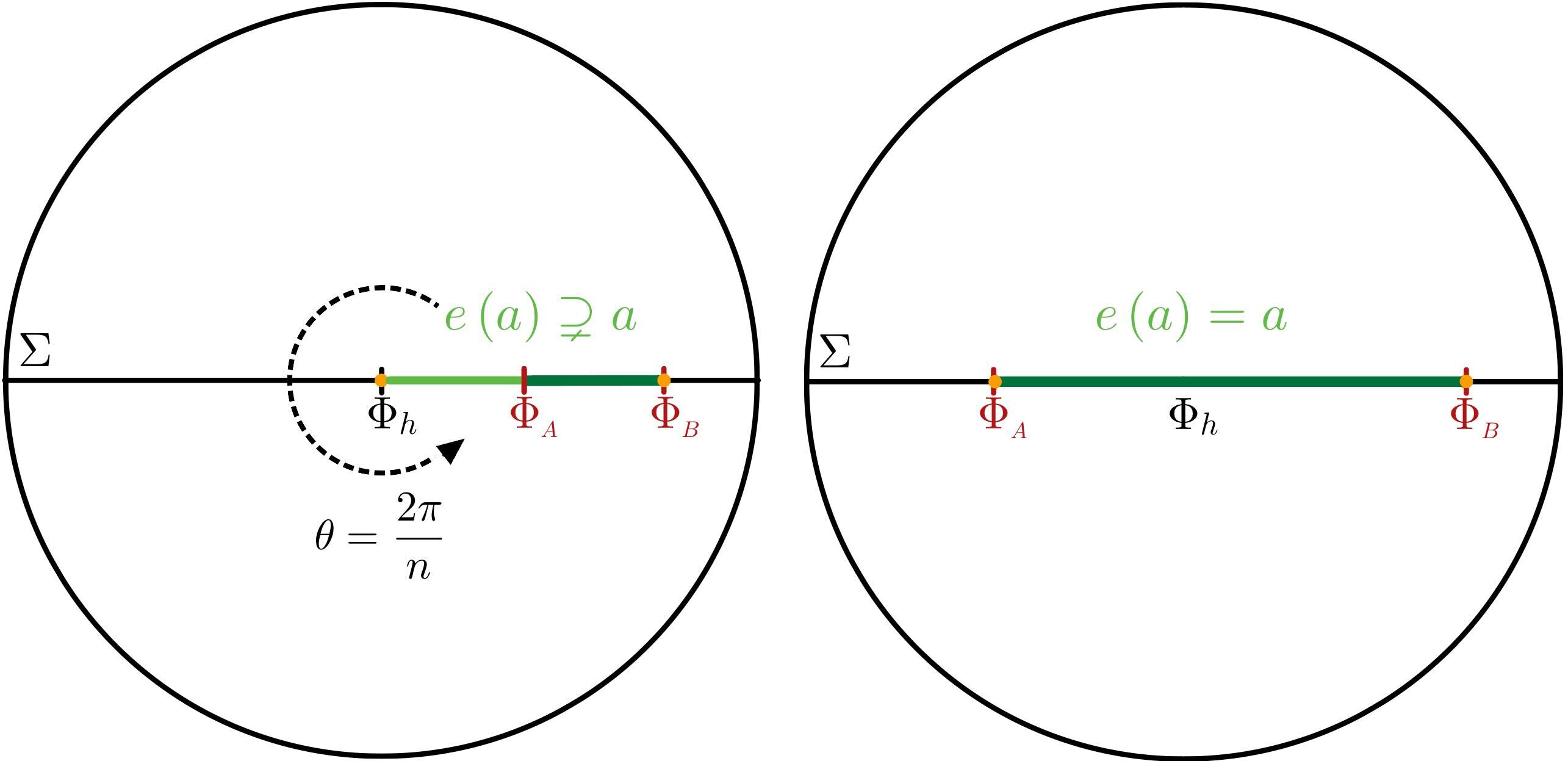}
    \caption{Left: For a bulk interval on a single side of the horizon, the entanglement wedge extends to the horizon. In the case where the value of the dilaton at the endpoints is specified, the opening angle at the horizon is given by $\frac{2\pi}{n}$ so that the parent spacetime is smooth. Right: For a bulk interval which crosses the horizon, the entanglement wedge is the region $a$ itself.}
    \label{fig:JT_wedges}
\end{figure}
Due to the radial monotonicity of $\Phi(x)$, in case I, the entanglement wedge of $a$ extends to the horizon, and properly contains $a$, i.e., $e(a)\supsetneq a $ while in case II, they are equal, i.e., $e(a)=a$. See \figref{fig:JT_wedges}. 

In what follows we consider case I, and present the result for case II in Appendix~\ref{sec:App_OppSide}. In any given basis state labelled by $\Phi_h$, the extremal surface homologous to $a$ is given by $\gamma_a = H\cup B$ where $H$ denotes the bifurcation point. Its area functional is
\begin{align}\label{eq:entropy}
    \frac{\mathcal{A}(\gamma_a,h)}{4G} = 2\pi\Phi_0+ 2\pi(\Phi_h +\Phi_B),
\end{align}
where we note that $2\pi\Phi$ plays the role of $\frac{\mathcal A}{4G}$ in JT gravity. Note that $\Phi_B$ in \Eqref{eq:entropy} is also implicitly a function of $\Phi_h$ depending on how we specify the subregion. With this setup in mind, we will now use a few different methods of gauge-invariantly specifying $a\subset\Sigma$ not only in the dominant semiclassical geometry, but in all fixed-dilaton states.

\subsubsection*{Fixing the Dilaton} 

\begin{figure}
    \centering
    \includegraphics[width=.9\linewidth]{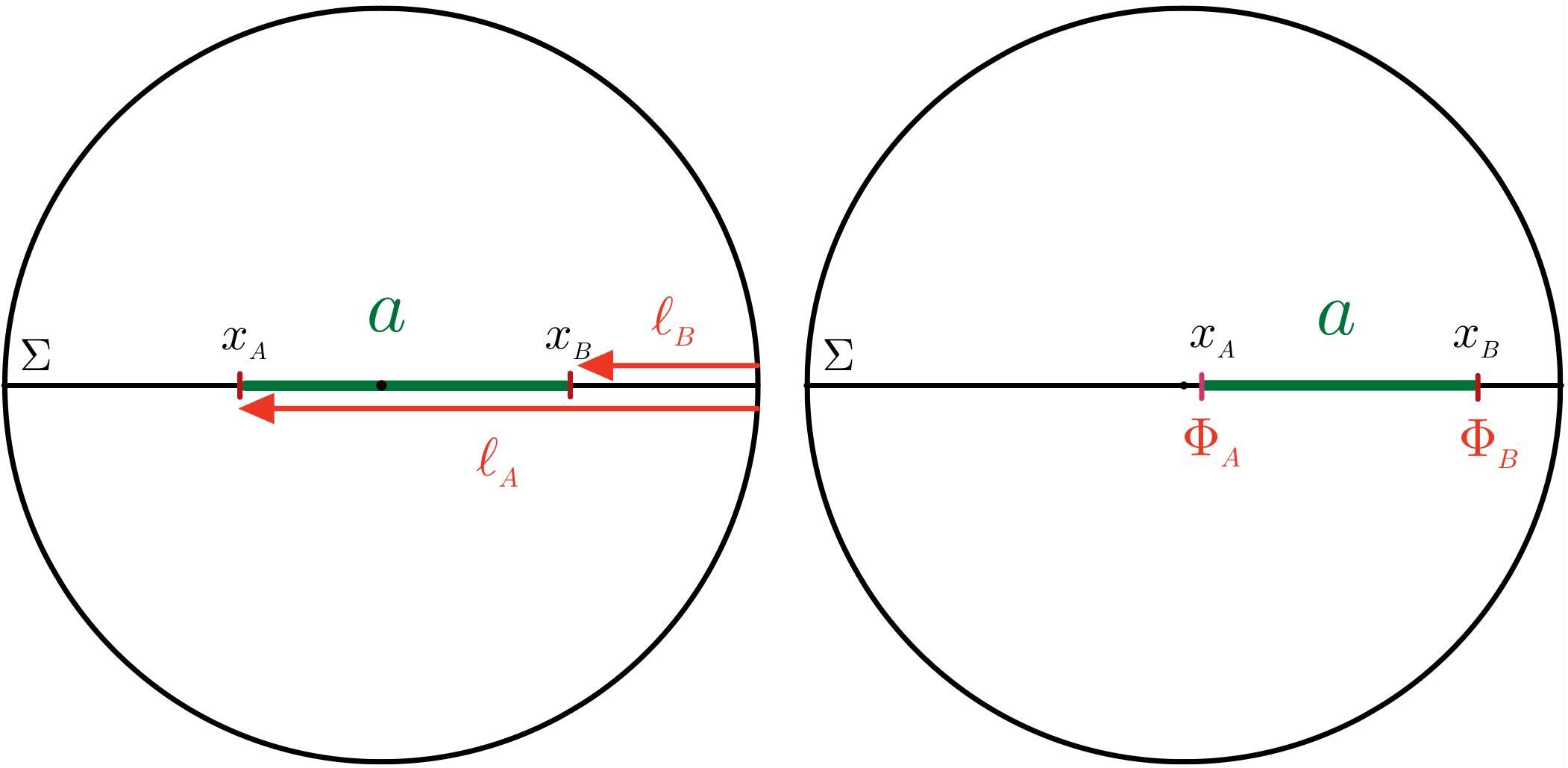}
    \caption{Two gauge-invariant specifications of region $a$: On the left, we specify the value of the dilaton at its endpoints, along with whether the points lie on the same or opposite sides of the horizon from a chosen asymptotic boundary. On the right, we specify the geodesic distance of its endpoints from a chosen asymptotic boundary.}
    \label{fig:JT_gauge_inv_a}
\end{figure}

The first and simplest gauge-invariant specification for $a\in\Sigma$ is to directly fix the values of the dilaton at its endpoints, along with which side of the horizon the endpoints lie. Suppose that $\Phi=\Phi_A$ and $\Phi=\Phi_B$ are both fixed on the right with $\Phi_B>\Phi_A>0$. See \figref{fig:JT_gauge_inv_a}.

To compute the R\'enyi entropy for the bulk region $a$, we substitute our expression for $\mathcal{A}(\gamma_a,h)$ into our master formula \Eqref{eq:Zma} to obtain\footnote{We must restrict to geometries where $\Phi_h\leq\Phi_A$, so that the subregion $a$ is well-defined. }
 \begin{align}\label{eq:JTRenyi}
     \text{Tr}\left(\rho_a^n\right) &= \sum_{h}|\psi(\Phi_h)|^{2n} \exp\{-(n-1)\frac{\mathcal{A}(\gamma_a,h)}{4G}\}\\
     &= \sum\limits_{\Phi_h\leq\Phi_A}\exp\{2\pi n(\Phi_0 + \Phi_h) - \frac{n\beta}{2}\Phi_h^2 - (n-1)2\pi(\Phi_0+\Phi_h+\Phi_B) \}
 \end{align}
Upon extremizing over values of $\Phi_h$, the saddle-point geometry $h_n$ is given by 
\begin{equation}\Phi_{h_n}= \frac{2\pi}{n\beta}
\end{equation}
Using \Eqref{eq:renyi_a}, we have
\begin{align}
    S_n(a) &=2\pi\Phi_0+ 2\pi(\frac{2\pi}{\beta}+\Phi_B) + \frac{1-n}{n}\frac{2\pi^2}{\beta}\\
    S(a) &= \lim\limits_{n\rightarrow 1}S_n(a) = 2\pi\Phi_0+2\pi(\frac{2\pi}{\beta}+\Phi_B)= \frac{\mathcal{A}(\gamma_a,h_\psi)}{4G},
\end{align}
where the last line is consistent with the BP proposal.

We note that the saddle-point in this example is related to the semiclassical geometry via $\beta\rightarrow n\beta$. This indicates an opening angle of $\theta=2\pi /n$. This is interesting since it demonstrates that the full spacetime is smooth once we glue together $n$ copies around the horizon. This occurs since specifying the dilaton value at the endpoints is a local specification and thus, no singularities are introduced in regions outside $a$. This will not be the case generically as we show in other examples.

\subsubsection*{Fixing the proper distance from the asymptotic boundary}

We now consider another method for gauge-invariantly specifying subregions $a$. Rather than directly fixing the value of the dilaton on $\partial a$, we fix the geodesic distance between $\partial a$ and an asymptotic boundary. The geodesic length $L$ from the boundary (say, at cutoff $x_c$) to a point $x_0$ is $L = \int_{x_0}^{x_c}\frac{1}{\sqrt{1+x^2}}dx=\sinh^{-1}x_c-\sinh^{-1}x_0$. This diverges as $x_c\rightarrow\infty$. Therefore, we define a renormalized geodesic distance as: 

\begin{equation}
    \ell \equiv L - \log(2\Phi|_{\partial\Sigma})
\end{equation}
where $\Phi|_{\partial\Sigma}=\Phi_h\sqrt{1+x_c^2}$. This gives

\begin{align}
    \ell\xrightarrow[x_c\rightarrow\infty]{}&=-\sinh^{-1}x_0-\log(2\Phi_h) = -\log\bigg(\Phi_h\Big(x_0 + \sqrt{1+x_0^2}\Big)\bigg)\\
    &= -\log\Big(\sqrt{\Phi(x_0)^2 - \Phi_h^2} + \Phi(x_0)\Big)
\end{align}
or equivalently
\begin{align}
    x_0 = -\sinh\Big(\ell + \log \Phi_h \Big)
\end{align}
Moreover, the value of the dilaton at $x_0$ is determined by $l$ and $\Phi_h$ as 
\begin{equation}
     \Phi(x_0) = \Phi_h\cosh\Big(\ell + \log \Phi_h\Big)
\end{equation}

Therefore, we can uniquely determine the location of $\partial a=A\cup B$ by specifying the renormalized geodesic distances $\ell_A$,$\ell_B$ of the endpoints from a chosen asymptotic boundary. See \figref{fig:JT_gauge_inv_a}. We assume without loss of generality that $\ell_A<\ell_B$. With these fixed, the extremal area becomes 
\begin{equation}
    \frac{\mathcal{A}(\gamma_a,h)}{4G} = 2\pi\Phi_0+2\pi(\Phi_h+\Phi_B) =2\pi\Phi_0+ 2\pi (\Phi_h+\Phi_h\cosh\left(\ell_B+\log\Phi_h\right)).
\end{equation}
The saddles for $\text{Tr}\left(\rho_a^n\right)$ are given by
\begin{align}
    \Phi_{h_n}&=\frac{2\pi/\beta}{\frac{2\pi}{\beta} e^{\ell_B}(n-1) + n}=\frac{2\pi}{\beta} - \frac{\pi}{\beta}\left(\frac{2\pi}{\beta}e^{\ell_B}+1\right)(n-1)+O((n-1)^2),
\end{align}
where we have expanded the exact formula near $n=1$. The corresponding entanglement entropies can then be checked to be
\begin{align}
    S(a)&=2\pi\Phi_0+2\pi(\Phi_{h_\psi}+\Phi_B) = \frac{\mathcal{A}(\gamma_a,h_\psi)}{4G} 
\end{align}
in agreement with the BP proposal as expected. 

An important point in the above calculation is that the parent spacetime obtained by gluing together $n$ copies will not be smooth at the horizon, unlike the previous example. This is allowed since the specification of the subregion is more non-local and thus, can backreact on the spacetime.

\subsubsection*{Fixing the dilaton relative to the horizon}
Instead of fixing the value of the dilaton at the boundary of the subregion, we can instead fix its value relative to the value at the horizon, in a given geometry. For example, we may specify $\Phi_A/\Phi_h = \phi_A$ and $\Phi_B/\Phi_h = \phi_B$. This again determines the location of the endpoints $x_A$ and $x_B$ up to which side of the geometry they lie on.\footnote{Unlike fixing the value of the dilaton directly, in this case we need not project onto the subspace of states with $\Phi_h<\Phi_A$ to have a well-defined subregion. Instead, for $\phi_A,\phi_B>0$, the subregion is guaranteed to exist for all choices of $\Phi_h$.} With $\phi_B>\phi_A>0$,  we have 
\begin{equation}
    \mathcal{A}(\gamma_a,h)=2\pi\Phi_0+2\pi\Phi_h(1+\phi_B)
\end{equation}
This gives the following saddle for $\text{Tr}\left(\rho_a^n\right)$:
 \begin{align}\Phi_{h_n}&= \frac{2\pi}{\beta} - \frac{(n-1)}{n}\frac{2\pi}{\beta}(1+\phi_B)
 \end{align}
The corresponding R\'enyi and Von Neumann entropies are 
\begin{align}
    S_n(a) &=2\pi\Phi_0+ \frac{4\pi^2}{\beta}(1+\phi_B) -\left(\frac{n-1}{n}\right)\frac{2\pi^2}{\beta}(1+\phi_B)^2\\
    S(a) &= 2\pi\Phi_0+\frac{4\pi^2}{\beta}(1+\phi_B) = \frac{\mathcal{A}(\gamma_a,h_\psi)}{4G}.
\end{align}

\subsubsection*{Fixing the proper distance from the horizon}

Instead of fixing the proper distance between $\partial a$ and the asymptotic boundary, we can also fix its distance from the horizon. Since this is a finite quantity, we need not use the renormalized geodesic distance, and we can fix the proper distance directly. Since proper distance is symmetric across the horizon, we must also specify, as before, which side of the geometry the endpoints lie on. To illustrate this, note that the distance from the horizon to a point $x_0$ is
\begin{equation}
    L = \int_{0}^{x_0}\frac{1}{\sqrt{1+x^2}}|dx| = \sinh^{-1}|x_0|
\end{equation}
Thus $x_0 = \pm\sinh(L)$. The value of the dilaton at $x_0$ is determined from $L$ and $\phi_h$ as
\begin{equation}
    \Phi(x_0)=\Phi_h\cosh L
\end{equation}
Without loss of generality, suppose $L_B>L_A$. Then,
\begin{align}
    \frac{\mathcal{A}(\gamma_a,h)}{4G} &= 2\pi\Phi_0+2\pi(\Phi_h + \Phi_B)=2\pi\Phi_0+ 2\pi\Phi_h(1+\cosh L_B)
\end{align}
The saddles of $\text{Tr}\left(\rho_a^n\right)$ occur at 
 \begin{equation}\Phi_{h_n}= \frac{2\pi}{\beta} - \frac{(n-1)}{n}\frac{2\pi}{\beta}(1+\cosh L_B)
 \end{equation}
which gives the R\'enyi and Von Neumann entropies
\begin{align}
    S_n(a) &= 2\pi\Phi_0+\frac{4\pi^2}{\beta}(1+\cosh L_B) -\left(\frac{n-1}{n}\right)\frac{2\pi^2}{\beta}(1+\cosh L_B)^2\\
    S(a) &= 2\pi\Phi_0+\frac{4\pi^2}{\beta}(1+\cosh l_B) = \frac{\mathcal{A}(\gamma_a,h_\psi)}{4G}
\end{align}
Again, the last line is in agreement with the BP proposal.

\subsection{Einstein Gravity : Minisuperspace Approximation} \label{sub:ein}

Similar to the calculation in JT gravity, we can construct the Hartle-Hawking state for $(d+1)$-dimensional black holes in the minisuperspace approximation to Einstein gravity where we assume spherical symmetry \cite{Chua:2023srl}. The state is likewise prepared via the Euclidean gravitational path integral with boundary conditions consisting of an asymptotic AdS boundary of length $\beta/2$, and a bulk timeslice with fixed geometry $h$. In the minisuperspace approximation, the geometry of slice $\Sigma$ can be uniquely determined by its energy, or equivalently by its horizon area. We find that in the basis of states with fixed horizon area, denoted $\ket{\mathcal{A}_h}$, the wavefunction is given by 
\begin{equation}
    \psi_{\beta/2}(\mathcal{A}_h) = \braket{\mathcal{A}_h|\psi_{\beta/2}}=e^{-I[g_h]}=\exp\{\frac{1}{2}S(\mathcal{A}_h) - \frac{\beta}{2}E(\mathcal{A}_h)\}
\end{equation}
where $S(\mathcal{A}_h) = \frac{\mathcal{A}_h}{4G}$ is the entropy, and $E(\mathcal{A}_h)$ is the energy of the slice $\Sigma$ in geometry $h$, determined through 

\begin{equation}
    E = \frac{(d-1)\Omega_{d-1}}{16\pi G_{N}}r_h^{d-2}(1+r_h^2)
\end{equation} 
where $r_h$ is the horizon radius, e.g. $\mathcal{A}_h=\Omega_{d-1}r_h^{d-1}$. See appendix \ref{sec:App_HH} for the details.

\begin{figure}
    \centering
    \includegraphics[width=.9\linewidth]{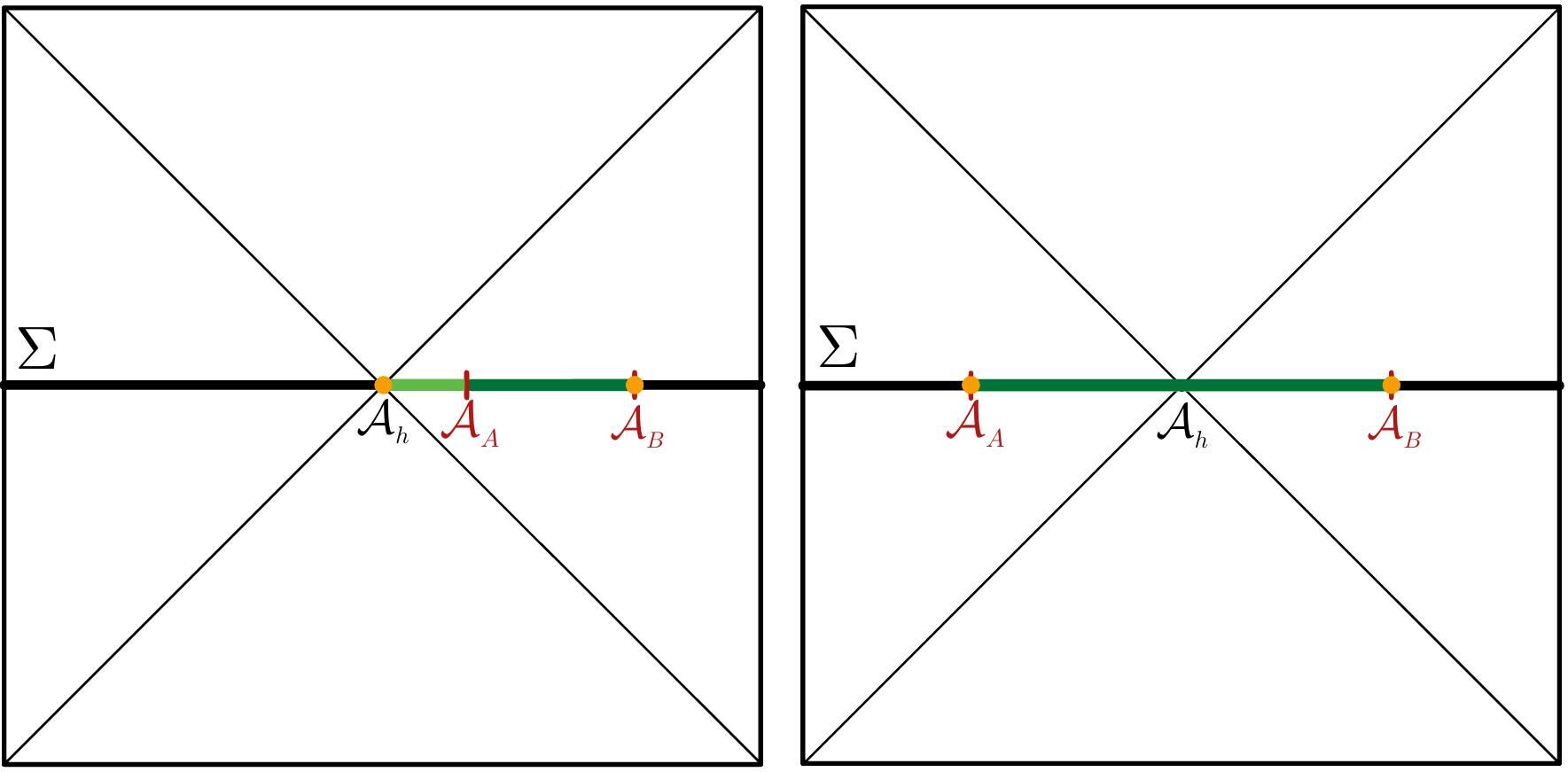}
    \caption{The region we choose is an annulus bounded by spheres with areas $\mathcal{A}_A$ and $\mathcal{A}_B$. On the left, the annulus is on the same exterior region in which case, the entanglement wedge extends out to the horizon. On the right, the annulus includes the horizon in which case, the entanglement wedge is the same as the region itself.}
    \label{Einstein_wedges}
\end{figure}

We consider simply-connected regions which are bounded by two spheres, say $A$ and $B$, which are centered around the origin. That is, $\partial a = A\cup B$. As in the case of dilaton gravity, we generically have two cases, where $A$ and $B$ bound a region on a single side of the geometry, or a region spanning the horizon, see \figref{Einstein_wedges}. We again leave the latter for appendix \ref{sec:App_OppSide}. Since the areas of spheres monotonically increase with their radius, in any geometry $h$, the extremal surface homologous to $\partial a$ is
\begin{align}
    \gamma_a &= H\cup B
\end{align}
where $A$ and $B$ are no longer spacetime points, but are $d-1$ dimensional spheres in $\Sigma$ centered around the horizon $H$. See \figref{Einstein_wedges}. Its area in geometry $h$ is
\begin{align}
    \mathcal{A}(\gamma_a,h) &= \mathcal{A}(H,h)+ \mathcal{A}(B,h)
\end{align}
The areas of these surfaces depend on how we gauge-invariantly specify the region $\partial a$. Analogous to fixing the value of the dilaton in JT gravity, the simplest specification is to directly fix the area of its bounding surfaces, along with which side of the geometry $A$ and $B$ lie on. Specifically, we fix $\mathcal{A}(A,h) \equiv \mathcal{A}_A$ and $\mathcal{A}(B,h) \equiv \mathcal{A}_B$, and specify whether $A$ and $B$ lie on the same or opposite sides of chosen asymptotic boundary. Note that $\mathcal{A}(H,h)=\mathcal{A}_h$ by definition. With this choice of specification, the R\'enyi entropies are computed as follows.
 \begin{equation}
     \text{Tr}\left(\rho_a^n\right) = \sum_{\mathcal{A}_h}|\psi(\mathcal{A}_h)|^{2n}\exp\{\frac{1}{2}S(\mathcal{A}_h) - \frac{\beta}{2}E(\mathcal{A}_h)-(n-1)\frac{\mathcal{A}(\gamma_a,h)}{4G}\}
     \end{equation}
The saddle-point value of the horizon area $\mathcal{A}_{h_n}$ is given by 
\begin{align}
    \mathcal{A}_{h_n} &= \Omega_{d-1}\left(\frac{2\pi}{d n\beta }+ \sqrt{\frac{4\pi^2}{d^2n^2\beta^2}-\(\frac{d-2}{d}\)}\right)^{d-1}
\end{align}
The resulting R\'enyi and Von Neumann entropies are 
\begin{align}
    S_n(a)&= \frac{1}{1-n}\log\left(\frac{\text{Tr}\left(\rho_a^n\right)}{(\text{Tr}\rho_a)^n}\right)=\frac{\mathcal{A}_{h_n} + \mathcal{A}_B}{4G} + O\left(1-n\right)\\
    S(a)&=\frac{\mathcal{A}_{h_\psi}+\mathcal{A}_B}{4G}=\frac{\mathcal{A}(\gamma_a,h_\psi)}{4G}
\end{align}
in agreement with the BP proposal. 

Finally, we also observe that the saddle for $\braket{\psi|\psi}$, namely $\mathcal{A}_\psi$, is related to $\mathcal{A}_{h_n}$ by $\beta\rightarrow n\beta$, indicating an opening angle of $\theta=\frac{2\pi}{n}$ around the horizon, as expected from the local specification of the subregion. This is a generalization of the standard Lewkowycz-Maldacena saddle where there is a conical defect of opening angle $2\pi n$ at $B$, but the geometry is smooth everywhere else.

\section{Discussion}
\label{sec:disc}
\subsection{Min and Max Entanglement Wedges}
\label{sub:min}
In this paper, we restricted our attention to semiclassical states for which entropies can be computed under the assumption of replica symmetry. However, Ref.~\cite{Akers:2020pmf} showed that this assumption breaks down for a class of states known as \emph{incompressible states}, which can be constructed straightforwardly even in static spacetimes. In such cases, one must invoke tools from one-shot state merging protocols. This framework requires distinguishing between \emph{max entanglement wedges}, where the probability of successful reconstruction is close to $1$, and \emph{min entanglement wedges}, where reconstruction is greater than some infinitesimal but nonzero $\epsilon$. Here, we briefly comment on how such min and max entanglement wedges may be obtained via the Petz map reconstruction.

In Ref.~\cite{Penington:2019kki}, the Petz map was explicitly computed using the gravitational path integral to determine operator reconstructability.\footnote{A similar calculation was done for RTNs in Ref.~\cite{Jia:2020etj}.} The authors identified two distinct regimes: one in which an operator is fully reconstructible and another in which it is entirely unreconstructible. In particular, they showed that the matrix elements of the boundary operator $O_R$ implements the following in the bulk:
\begin{equation}
    \bra{a}O_R\ket{b} = c_1 \bra{a}O\ket{b} + c_2 \delta_{ab} \mathrm{Tr}\, O ~,
\end{equation}
where $\bra{a}O\ket{b}$ is the matrix elements of the operator in the bulk region. They estimated the coefficients $c_1$ and $c_2$ to quantify the quality of reconstruction.

Although Ref.~\cite{Penington:2019kki} found a sharp transition between full reconstructability and non-reconstructability, this transition can be made more gradual by increasing the code space dimension. This allows for intermediate regimes in which reconstruction is possible with a finite $O(1)$ probability. In principle, for a fixed input bulk region, one could similarly examine reconstructability of operators localized at different points in spacetime by computing the Petz map and estimating the coefficients $c_1$ and $c_2$. The reconstruction success probability is then given by $1 - c_2$. If this probability is sufficiently high, i.e., for $0<c_2<\epsilon$, the operator can be considered to be part of the max entanglement wedge. On the other hand, if this probability is low but non-zero, i.e., for $\epsilon<c_2<1$, it belongs to the min wedge.

While this method is not practical for general computations, it provides an \emph{in principle} path integral prescription for determining min and max entanglement wedges.

\subsection{Time Dependence}
\label{sub:time}

An important direction for future work is the open problem of extending our methods to time-dependent settings. Our primary inspiration, the tensor network picture, does not straightforwardly generalize beyond static space-times. This limitation is compounded by the fact that the distinction between min and max entanglement wedges already arises at the classical level. This suggests that the path integral construction used to compute entanglement wedges in general time-dependent spacetimes must go beyond the standard assumption of replica symmetry. Although Ref.~\cite{Akers:2020pmf} demonstrated how to identify min and max entanglement wedges of boundary subregions in static settings using fixed-area states, it remains poorly understood how to define or compute such wedges in time-dependent geometries, even for boundary input regions. Despite these challenges, progress can still be made in certain restricted scenarios. We will explore some of these developments in forthcoming work~\cite{kaya2025}.

\begin{figure}
    \centering
    \includegraphics[width=0.5\linewidth]{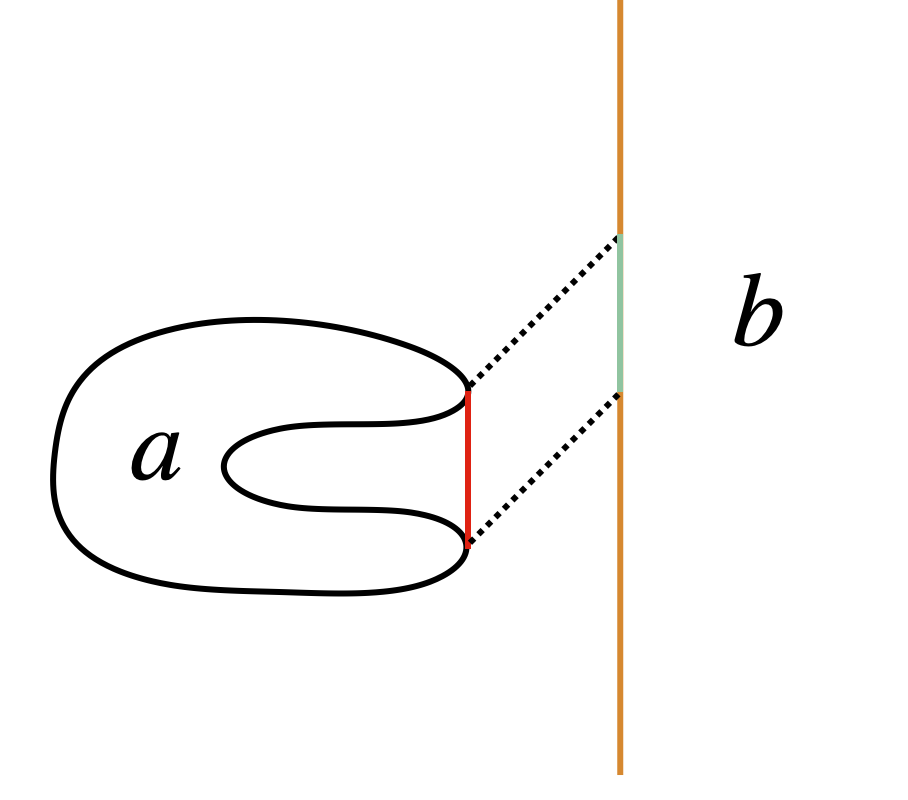}
    \caption{
    An example configuration illustrating subtleties that arise in time-dependent settings. For simplicity, we work at leading order in $G$ and choose wedges $a$ and $b$ such that we have $\emax = \emin = e$. In this case, $e(a)$ is given by a restricted maximin procedure (see Ref.~\cite{kaya2025} for a proof of this statement).
Consider a subregion $a$ that forms a horseshoe-shaped wedge on a Cauchy slice in asymptotically flat spacetime. The portion of the boundary of $e(a)$ that does not coincide with the boundary of $a$ is shown in red. Subregion $b$ is chosen such that a portion of its boundary (shown in green) is null like to the red segment of $e(a)$. Generically, this green portion of $\partial b$ will have smaller area than the red portion of $e(a)$. We also choose $b$ such that it is normal everywhere such that $e(b)=b$. Thus $a$ and $b$ satisfy $a \subset e(b)'$ and $b \subset e(a)'$.
Since computing the restricted entanglement wedge $e(a)|_{b'}$ requires all candidate Cauchy slices to include the boundary of $b$, the surface $e(a)$ can no longer be a valid maximin surface in $b'$. Instead, a new surface—comprising the green portion of $\partial b$ along with the null surfaces that connect it to region $a$—will have smaller area and be present on all Cauchy slices. This surface therefore becomes $e(a)|_{b'}$.
    }
    \label{fig:td_indep_example}
\end{figure}

Apart from the gravitational path integral derivation, there are also questions to answer about entropy inequalities and independence conditions. An interesting subtlety that arises in the context of independence conditions is that, in time-dependent settings, an independence condition of the form given in \Eqref{eq:indep} is not sufficient to guarantee that the restricted entanglement wedges coincide with their unrestricted counterparts. For an explicit example, see \figref{fig:td_indep_example}. In this light, it may seem surprising that Ref.~\cite{Bousso:2024ysg} was able to use such an independence condition to prove the monogamy of mutual information (MMI). A more detailed investigation of independence conditions in time-dependent scenarios is an important direction for future work.

A refinement of the BP proposal was recently put forward by Ref.~\cite{Bousso:2025fgg}, which modifies the boundary conditions in both time-dependent and time-independent settings by introducing the fundamental complement of a bulk subregion. We note that the construction in Ref.~\cite{Bousso:2025fgg} is formulated entirely in Lorentzian signature. The essential features of this approach—including the role of the fundamental complement as a boundary condition—are expected to arise naturally within a fully Lorentzian path integral framework and are not anticipated to emerge in the Euclidean path integral setting employed in this work. We leave a detailed investigation of this direction for future work.

\acknowledgments

We thank Chris Akers, Raphael Bousso, Matthew Headrick, Don Marolf, Geoff Penington, Ronak Soni, Arvin Shahbazi-Moghaddam and Zhenbin Yang for useful discussions.
This work was supported in part by the Berkeley Center for Theoretical Physics; and
by the Department of Energy, Office of Science, Office of High Energy Physics under
Award DE-SC0025293.

\appendix

\section{The Hartle-Hawking Wavefunction}

\subsection{JT Gravity}

The action of interest is the Euclidean JT gravity action:\footnote{Note we remove a relative factor of 2 compared to Ref.~\cite{Harlow:2018tqv} to match the normalization used for JT gravity in other literature.}
\begin{align}\label{eq:JT_Action}
    I_E = -\frac{1}{2}\int\limits_M d^2x\sqrt{-g}\Big(\Phi_0 R + \Phi (R+2)\Big)  &-\int\limits_{\partial M}dt\sqrt{|\gamma|} \Big(\Phi_0K_{\partial\mathcal{M}} + \Phi(K_{\partial\mathcal{M}}-1)\Big)\\
    &-\int\limits_{\Sigma}dt\sqrt{|\gamma|} \Big(\Phi_0K_{\Sigma} + \Phi(K_{\Sigma}-1)\Big)
\end{align}
where $K_{\partial\mathcal{M}}$ and $K_\Sigma$ are the extrinsic curvatures on the AdS and bulk boundaries. We take $K_{\partial\mathcal{M}} = \gamma^{\mu\nu}\nabla_{\mu}r_\nu$ where $r^\mu$ is the normal vector in $\partial\mathcal{M},$ $\gamma^{\mu\nu}$ is the induced metric and $\Phi$ is the dilaton field. 
The asymptotic boundary conditions on $\partial\mathcal{M}$ are chosen to be
\begin{align*}
    \gamma_{t_Et_E}\big|_{\partial M} &= \frac{1}{r_c^2}\\ 
    \Phi \big|_{\partial M} &= r_c
\end{align*}
with $r_c\rightarrow\infty$. The boundary conditions on the timeslice $\Sigma$ are taken to be 
\begin{align}\label{eq:JTSlice1}
    n_\mu\partial^\mu\Phi_{\Sigma}=0\\ \label{eq:JTSlice2}
    \Phi_{\Sigma} = \Phi_h\sqrt{1+x^2}
\end{align}
Thus, $\Sigma$ is the time-reflection symmetric slice, and the value of the dilaton at the bifurcation point is $\Phi_h$. These conditions amount to projecting $\ket{\psi_{\beta/2}}$ onto the fixed-dilaton state $\ket{\Phi_h}$. 

The saddle point of the Euclidean action is simply the section of the Euclidean Schwarzschild solution with boundary length $r_c\beta /2$, i.e. with $t_E\in(0,\beta/2)$. See \figref{fig:JT_psi_h}. Therefore, the saddle-point geometry covers only a fraction of the entire black hole geometry and has an opening angle determined by the ratio $\frac{\beta/2}{\beta_h}$. In particular, the opening angle is $\theta = \pi\frac{\beta}{\beta_h} = \frac{\beta}{2}\Phi_h$.

In Schwarzschild coordinates, the horizon is located at $r_h=\Phi_h$. Since we do \textit{not} impose smoothness at the horizon, $\Phi_h\neq 2\pi /\beta$. Rather, there is a conical defect at the horizon of angle $\theta = \beta\Phi_h/2$. 
The contribution to the action from this corner\footnote{There are also corners of angle $\pi/2$ where the AdS boundary meets the bulk slice $\Sigma$, but these have a vanishing contribution to the action. } is  

\begin{equation*}
    I_{\text{corner}} = -(\Phi_0 + \Phi_h) (\pi - \theta)
\end{equation*}
The contribution away from the corners is simply a fraction of the action for the full (smooth) geometry: 
\begin{equation}
    I[g_h]=\frac{\beta/2}{2\pi/\Phi_h}I_{tot}
\end{equation}
Evaluating the Euclidean action including the defect contribution gives \cite{Harlow:2018tqv}
 \begin{equation}
     I_E[g_h] =-\pi(\Phi_0 + \Phi_h)+\frac{\beta}{4}\Phi_h^2 
 \end{equation}
and therefore 
\begin{equation}\label{eq:JTHH2}
\psi(\Phi_h)\approx e^{-I_E[g_h]}= \exp{\{\pi(\Phi_0 + \Phi_h)-\frac{\beta}{4}
 \Phi_h^2\}}
 \end{equation}

\subsection{Einstein Gravity}\label{sec:App_HH}
Here, we primarily follow the construction of Ref.~\cite{Chua:2023srl}. With a $d+1$ dimensional bulk $\mathcal{M}$, we impose the asymptotic AdS boundary conditions 
\begin{equation}
    \gamma_{ab}dy^ady^b = r_c^2 (d\tau^2 + d\Omega_{d-1}^2)
\end{equation}
where $r_c\rightarrow\infty$. The geometry of the bulk slice $\Sigma$ is taken to be that of an eternal black hole with fixed horizon area $\mathcal{A}_h$, or equivalently, a fixed energy $E(\mathcal{A}_h)$\footnote{In this way, we may identify states $\ket{E}$ of energy with states of fixed horizon area $\ket{A_h}$, analogous to the fixed dilaton states used in \secref{sec:eg}.}.  Specifically, 

\begin{equation}\label{Einstein_Sigma}
    h_{ij}dx^idx^j = d\rho^2 + r_h^2\cosh^2\rho d\Omega_{d-2}^2 
\end{equation}
where $\rho$ is the radial coordinate. The bulk slice $\Sigma$ meets the AdS boundary at $\rho_c=\pm\log\frac{2r_c}{r_h}$, and $r_h$ is the horizon radius, related to $E$ through 
\begin{equation}
    E(\mathcal{A}_h) = \frac{(d-1)\Omega_{d-1}}{16\pi G_{N}}r_h^{d-2}(1+r_h^2)
\end{equation} 
with $r_h$ defined through $\mathcal{A}_h=\Omega_{d-1}r_h^{d-1}$

The Euclidean action of interest is 

\begin{align}
    I_E = -\frac{1}{16\pi G}\int\limits_{\mathcal{M}}d^{d+1}x\sqrt{g}(R-2\Lambda) &-\frac{1}{8\pi G}\int\limits_{\partial\mathcal{M}}d^{d}y \sqrt{\gamma}K_{\partial\mathcal{M}}\\
    &-\frac{1}{8\pi G}\int\limits_{\Sigma}d^{d}y \sqrt{\gamma}K_{\Sigma}
\end{align}
where $K_{\partial\mathcal{M}}$ and $K_\Sigma$ are the extrinsic curvatures on the asymptotic and bulk boundaries. 
The unnormalized Hartle-Hawking wavefunction $\psi_{\beta/2}$ is again given simply by the path integral in \Eqref{eq:path} 
\begin{equation}
    \psi_{\beta/2}(E) \approx e^{-I_E[g_h]},
\end{equation}
where $g_h$ is the saddle point of the action with fixed geometry $h$ on $\Sigma$, corresponding to horizon radius $r_h$. The saddle-point geometry, as in JT gravity, is given by a slice of the eternal black hole geometry with an asymptotic boundary of length $\beta/2$, with a conical defect at the horizon (as in \figref{fig:JT_psi_h}). At fixed energy $E$, the black hole geometry has an inverse temperature 
\begin{equation}
    \beta_h = 4\pi \frac{r_h}{r_h^2d + d-2}
\end{equation}
Therefore the saddle geometry covers only a fraction of the entire black hole geometry, and has opening angle determined by the ratio $\frac{\beta/2}{\beta_h}$. In particular, the opening angle is $\theta = \pi\frac{\beta}{\beta_h}$.

The saddle-point action has two contributions, coming from the fraction of the full black hole geometry, and from the corners at the horizon, and where $\Sigma$ meets $\partial \mathcal{M}$. The full black hole geometry has action
\begin{equation}
    I_{tot}[g_h] = \beta_h E(\mathcal{A}_h) - S(\mathcal{A}_h)
\end{equation}
So, including the contribution from the corners, the action of the saddlepoint geometry evaluates to 

\begin{equation}
    I[g_h] = \frac{\beta/2}{\beta_h} I_{tot}[g_h] + I_{\text{corners}} = -\frac{1}{2}S(\mathcal{A}_h) + \frac{\beta}{2}E(\mathcal{A}_h)
\end{equation}.
This gives 
\begin{equation}
    \psi_{\beta/2}(\mathcal{A}_h) = \exp\{\frac{1}{2}S(\mathcal{A}_h) - \frac{\beta}{2}E(\mathcal{A}_h)\}
\end{equation}

\section{Bulk Subregions Spanning the Horizon}\label{sec:App_OppSide}
We briefly state the results for the R\'enyi entropies of bulk regions which span the horizon, using our four different gauge-invariant specifications. Specifically, we consider the case where the endpoints $A$ and $B$ lie on opposite sides of the horizon, so that $e(a)=a$, and hence $ \gamma_a = A\cup B$ with area
\begin{equation}
\mathcal{A}(\gamma_a,h) = \mathcal{A}(A,h) + \mathcal{A}(B,h)
\end{equation}
where $\mathcal{A}(A,h)$ and $\mathcal{A}(B,h)$ are determined by our gauge-invariant subregion specification. 

\subsection*{Fixed Dilaton Value}
In this case, we have $\frac{\mathcal{A}(\gamma_a,h)}{4G} = 2\pi\Phi_0+2\pi (\Phi_A + \Phi_B)$ which gives 
 \begin{align}
     \Phi_{h_n} &= \frac{2\pi}{\beta}\\
    S_n(a) &= 2\pi\Phi_0+2\pi(\Phi_A+\Phi_B)=\frac{\mathcal{A}(\gamma_a,h_\psi)}{4G}\\
    S(a) &= 2\pi\Phi_0+2\pi(\Phi_A+\Phi_B) = \frac{\mathcal{A}(\gamma_a,h_\psi)}{4G}
\end{align}

\subsection*{Fixed Renormalized Distance}
In this case, we have \begin{equation}
    \frac{\mathcal{A}(\gamma_a,h)}{4G} =2\pi\Phi_0+2\pi \Phi_h\left(\cosh(\ell_A+\log\Phi_h) + \cosh(\ell_B + \log\Phi_h)\right)
\end{equation} which gives 
\begin{align}
    \Phi_{h_n}&=\frac{2\pi/\beta}{\frac{2\pi}{n\beta}(e^{\ell_A}+e^{\ell_B})(m-1) +1} = \frac{2\pi}{\beta} -\frac{4\pi^2}{\beta^2}(e^{\ell_A}+e^{\ell_B})(n-1)+O((n-1)^2)\\
     S_n(a) &= 2\pi\Phi_0+2\pi\frac{2\pi}{\beta}\cosh({\ell_A} + \log\frac{2\pi}{\beta})+ 2\pi\frac{2\pi}{\beta}\cosh({\ell_B} + \log\frac{2\pi}{\beta})\ + O\left(n-1\right)\\
    S(a) &= 2\pi\Phi_0+2\pi(\Phi_A+\Phi_B)=\frac{\mathcal{A}(\gamma_a,h_\psi)}{4G}
\end{align}

\subsection*{Fixed Relative Dilaton Value}
In this case, $\mathcal{A}(\gamma_a,h) = 2\pi\Phi_0+2\pi\Phi_h(\phi_A+\phi_B)$ which gives
\begin{align}
    \Phi_{h_n}&= \frac{2\pi}{\beta} - \frac{(n-1)}{n}\frac{\pi}{\beta}(\phi_A+\phi_B)\\
    S_n(a) &= 2\pi\Phi_0+\frac{4\pi^2}{\beta}(\phi_A+\phi_B) -\left(\frac{n-1}{n}\right)\frac{2\pi^2}{\beta}(\phi_A+\phi_B)^2\\
    S(a) &= 2\pi\Phi_0+\frac{4\pi^2}{\beta}(\phi_A+\phi_B) = \frac{\mathcal{A}(\gamma_a,h_\psi)}{4G}
\end{align}

\subsection*{Fixed Distance from Horizon}
With $\frac{\mathcal{A}(\gamma_a,h)}{4G} = 2\pi\Phi_0+ 2\pi \Phi_h(\cosh L_A+\cosh L_B)$, we have 

\begin{align}
    \Phi_{h_n}&= \frac{2\pi}{\beta} - \frac{(n-1)}{n}\frac{2\pi}{\beta}(\cosh L_A+\cosh L_B)\\
     S_n(a) &= 2\pi\Phi_0+\frac{4\pi^2}{\beta}(\cosh L_A+\cosh L_B) -\left(\frac{n-1}{n}\right)\frac{2\pi^2}{\beta}(\cosh L_A+\cosh L_B)^2\\ 
    S(a) &=2\pi\Phi_0+\frac{4\pi^2}{\beta}
    (\cosh L_A+\cosh L_B) = \frac{\mathcal{A}(\gamma_a,h_\psi)}{4G}
\end{align}

\subsection*{Fixed Horizon Area in Einstein Gravity} In this case, we have $\mathcal{A}(\gamma_a,h) = \mathcal{A}_A + \mathcal{A}_B$. This gives a saddlepoint geometry with horizon area 

\begin{equation}
  \mathcal{A}_{h_n} = \Omega_{d-1}\left(\frac{2\pi}{d\beta}+\sqrt{\frac{4\pi^2}{d^2\beta^2}-\(\frac{d-2}{d}\)}\right)^{d-1}
\end{equation}

Note that, as in the fixed-dilaton specification, the saddlepoint has trivial $n$ dependence, and agrees with the expected semiclassical horizon area $\mathcal{A}_{h_\psi}$. This arises from the fact that $\gamma_a$ is independent of $h$. Similarly, the spectrum is flat. The R\'enyi and Von Neumann entropies are 
\begin{align}
     S_n(a) &= \frac{\mathcal{A}_A+\mathcal{A}_B}{4G}=\frac{\mathcal{A}(\gamma_a,h_\psi)}{4G}\\
     S(a) &= \frac{\mathcal{A}_A+\mathcal{A}_B}{4G} = \frac{\mathcal{A}(\gamma_a,h_\psi)}{4G}
\end{align}

 \bibliographystyle{JHEP}
\bibliography{hollowgrams.bib}

\providecommand{\href}[2]{#2}\begingroup\raggedright\begin{thebibliography}{10}

\bibitem{Saad:2019lba}
P.~Saad, S.~H. Shenker, and D.~Stanford, {\it {JT gravity as a matrix integral}},  \href{http://arxiv.org/abs/1903.11115}{{\tt arXiv:1903.11115}}.

\bibitem{Stanford:2019vob}
D.~Stanford and E.~Witten, {\it {JT gravity and the ensembles of random matrix theory}},  {\em Adv. Theor. Math. Phys.} {\bf 24} (2020), no.~6 1475--1680, [\href{http://arxiv.org/abs/1907.03363}{{\tt arXiv:1907.03363}}].

\bibitem{Maxfield:2020ale}
H.~Maxfield and G.~J. Turiaci, {\it {The path integral of 3D gravity near extremality; or, JT gravity with defects as a matrix integral}},  {\em JHEP} {\bf 01} (2021) 118, [\href{http://arxiv.org/abs/2006.11317}{{\tt arXiv:2006.11317}}].

\bibitem{Witten:2020wvy}
E.~Witten, {\it {Matrix Models and Deformations of JT Gravity}},  {\em Proc. Roy. Soc. Lond. A} {\bf 476} (2020), no.~2244 20200582, [\href{http://arxiv.org/abs/2006.13414}{{\tt arXiv:2006.13414}}].

\bibitem{Saad:2018bqo}
P.~Saad, S.~H. Shenker, and D.~Stanford, {\it {A semiclassical ramp in SYK and in gravity}},  \href{http://arxiv.org/abs/1806.06840}{{\tt arXiv:1806.06840}}.

\bibitem{Penington:2019kki}
G.~Penington, S.~H. Shenker, D.~Stanford, and Z.~Yang, {\it {Replica wormholes and the black hole interior}},  \href{http://arxiv.org/abs/1911.11977}{{\tt arXiv:1911.11977}}.

\bibitem{Almheiri:2019qdq}
A.~Almheiri, T.~Hartman, J.~Maldacena, E.~Shaghoulian, and A.~Tajdini, {\it {Replica Wormholes and the Entropy of Hawking Radiation}},  {\em JHEP} {\bf 05} (2020) 013, [\href{http://arxiv.org/abs/1911.12333}{{\tt arXiv:1911.12333}}].

\bibitem{Geng:2021hlu}
H.~Geng, A.~Karch, C.~Perez-Pardavila, S.~Raju, L.~Randall, M.~Riojas, and S.~Shashi, {\it {Inconsistency of islands in theories with long-range gravity}},  {\em JHEP} {\bf 01} (2022) 182, [\href{http://arxiv.org/abs/2107.03390}{{\tt arXiv:2107.03390}}].

\bibitem{Raju:2021lwh}
S.~Raju, {\it {Failure of the split property in gravity and the information paradox}},  {\em Class. Quant. Grav.} {\bf 39} (2022), no.~6 064002, [\href{http://arxiv.org/abs/2110.05470}{{\tt arXiv:2110.05470}}].

\bibitem{Bousso:2022hlz}
R.~Bousso and G.~Penington, {\it {Entanglement wedges for gravitating regions}},  {\em Phys. Rev. D} {\bf 107} (2023), no.~8 086002, [\href{http://arxiv.org/abs/2208.04993}{{\tt arXiv:2208.04993}}].

\bibitem{Bousso:2023sya}
R.~Bousso and G.~Penington, {\it {Holograms in our world}},  {\em Phys. Rev. D} {\bf 108} (2023), no.~4 046007, [\href{http://arxiv.org/abs/2302.07892}{{\tt arXiv:2302.07892}}].

\bibitem{Dong:2020uxp}
X.~Dong, X.-L. Qi, Z.~Shangnan, and Z.~Yang, {\it {Effective entropy of quantum fields coupled with gravity}},  {\em JHEP} {\bf 10} (2020) 052, [\href{http://arxiv.org/abs/2007.02987}{{\tt arXiv:2007.02987}}].

\bibitem{Ryu:2006bv}
S.~Ryu and T.~Takayanagi, {\it {Holographic derivation of entanglement entropy from AdS/CFT}},  {\em Phys. Rev. Lett.} {\bf 96} (2006) 181602, [\href{http://arxiv.org/abs/hep-th/0603001}{{\tt hep-th/0603001}}].

\bibitem{Bousso:2024ysg}
R.~Bousso and S.~Kaya, {\it {Geometric quantum states beyond the AdS/CFT correspondence}},  {\em Phys. Rev. D} {\bf 110} (2024), no.~6 066017, [\href{http://arxiv.org/abs/2404.11644}{{\tt arXiv:2404.11644}}].

\bibitem{Bousso:2025mdp}
R.~Bousso and S.~Kaya, {\it {Holographic Entropy Cone Beyond AdS/CFT}},  \href{http://arxiv.org/abs/2502.03516}{{\tt arXiv:2502.03516}}.

\bibitem{Balasubramanian:2023dpj}
V.~Balasubramanian and C.~Cummings, {\it {The entropy of finite gravitating regions}},  \href{http://arxiv.org/abs/2312.08434}{{\tt arXiv:2312.08434}}.

\bibitem{Lin:2017uzr}
J.~Lin, {\it {Ryu-Takayanagi Area as an Entanglement Edge Term}},  \href{http://arxiv.org/abs/1704.07763}{{\tt arXiv:1704.07763}}.

\bibitem{Lin:2018xkj}
J.~Lin, {\it {Entanglement entropy in Jackiw-Teitelboim Gravity}},  \href{http://arxiv.org/abs/1807.06575}{{\tt arXiv:1807.06575}}.

\bibitem{Jafferis:2019wkd}
D.~L. Jafferis and D.~K. Kolchmeyer, {\it {Entanglement Entropy in Jackiw-Teitelboim Gravity}},  \href{http://arxiv.org/abs/1911.10663}{{\tt arXiv:1911.10663}}.

\bibitem{Hayden:2016cfa}
P.~Hayden, S.~Nezami, X.-L. Qi, N.~Thomas, M.~Walter, and Z.~Yang, {\it {Holographic duality from random tensor networks}},  {\em JHEP} {\bf 11} (2016) 009, [\href{http://arxiv.org/abs/1601.01694}{{\tt arXiv:1601.01694}}].

\bibitem{Dong:2021clv}
X.~Dong, X.-L. Qi, and M.~Walter, {\it {Holographic entanglement negativity and replica symmetry breaking}},  {\em JHEP} {\bf 06} (2021) 024, [\href{http://arxiv.org/abs/2101.11029}{{\tt arXiv:2101.11029}}].

\bibitem{Akers:2021pvd}
C.~Akers, T.~Faulkner, S.~Lin, and P.~Rath, {\it {Reflected entropy in random tensor networks}},  {\em JHEP} {\bf 05} (2022) 162, [\href{http://arxiv.org/abs/2112.09122}{{\tt arXiv:2112.09122}}].

\bibitem{Akers:2022zxr}
C.~Akers, T.~Faulkner, S.~Lin, and P.~Rath, {\it {Reflected entropy in random tensor networks. Part II. A topological index from canonical purification}},  {\em JHEP} {\bf 01} (2023) 067, [\href{http://arxiv.org/abs/2210.15006}{{\tt arXiv:2210.15006}}].

\bibitem{Akers:2023fqr}
C.~Akers, A.~Levine, G.~Penington, and E.~Wildenhain, {\it {One-shot holography}},  {\em SciPost Phys.} {\bf 16} (2024), no.~6 144, [\href{http://arxiv.org/abs/2307.13032}{{\tt arXiv:2307.13032}}].

\bibitem{Akers:2024pgq}
C.~Akers, T.~Faulkner, S.~Lin, and P.~Rath, {\it {Reflected entropy in random tensor networks. Part III. Triway cuts}},  {\em JHEP} {\bf 12} (2024) 209, [\href{http://arxiv.org/abs/2409.17218}{{\tt arXiv:2409.17218}}].

\bibitem{Akers:2018fow}
C.~Akers and P.~Rath, {\it {Holographic Renyi Entropy from Quantum Error Correction}},  {\em JHEP} {\bf 05} (2019) 052, [\href{http://arxiv.org/abs/1811.05171}{{\tt arXiv:1811.05171}}].

\bibitem{Dong:2018seb}
X.~Dong, D.~Harlow, and D.~Marolf, {\it {Flat entanglement spectra in fixed-area states of quantum gravity}},  {\em JHEP} {\bf 10} (2019) 240, [\href{http://arxiv.org/abs/1811.05382}{{\tt arXiv:1811.05382}}].

\bibitem{Dong:2019piw}
X.~Dong and D.~Marolf, {\it {One-loop universality of holographic codes}},  {\em JHEP} {\bf 03} (2020) 191, [\href{http://arxiv.org/abs/1910.06329}{{\tt arXiv:1910.06329}}].

\bibitem{Penington:2022dhr}
G.~Penington, M.~Walter, and F.~Witteveen, {\it {Fun with replicas: tripartitions in tensor networks and gravity}},  {\em JHEP} {\bf 05} (2023) 008, [\href{http://arxiv.org/abs/2211.16045}{{\tt arXiv:2211.16045}}].

\bibitem{Marolf:2020vsi}
D.~Marolf, S.~Wang, and Z.~Wang, {\it {Probing phase transitions of holographic entanglement entropy with fixed area states}},  {\em JHEP} {\bf 12} (2020) 084, [\href{http://arxiv.org/abs/2006.10089}{{\tt arXiv:2006.10089}}].

\bibitem{Akers:2020pmf}
C.~Akers and G.~Penington, {\it {Leading order corrections to the quantum extremal surface prescription}},  {\em JHEP} {\bf 04} (2021) 062, [\href{http://arxiv.org/abs/2008.03319}{{\tt arXiv:2008.03319}}].

\bibitem{Dong:2023bfy}
X.~Dong, J.~Kudler-Flam, and P.~Rath, {\it {A modified cosmic brane proposal for holographic Renyi entropy}},  {\em JHEP} {\bf 06} (2024) 120, [\href{http://arxiv.org/abs/2312.04625}{{\tt arXiv:2312.04625}}].

\bibitem{Penington:2024jmt}
G.~Penington and P.~Rath, {\it {Diagonal Approximation for Holographic R\'enyi Entropies}},  {\em Phys. Rev. Lett.} {\bf 134} (2025), no.~16 161501, [\href{http://arxiv.org/abs/2412.03670}{{\tt arXiv:2412.03670}}].

\bibitem{Lewkowycz:2013nqa}
A.~Lewkowycz and J.~Maldacena, {\it {Generalized gravitational entropy}},  {\em JHEP} {\bf 08} (2013) 090, [\href{http://arxiv.org/abs/1304.4926}{{\tt arXiv:1304.4926}}].

\bibitem{Dong:2016fnf}
X.~Dong, {\it {The Gravity Dual of Renyi Entropy}},  {\em Nature Commun.} {\bf 7} (2016) 12472, [\href{http://arxiv.org/abs/1601.06788}{{\tt arXiv:1601.06788}}].

\bibitem{Dong_2024}
X.~Dong, J.~Kudler-Flam, and P.~Rath, {\it A modified cosmic brane proposal for holographic renyi entropy},  {\em Journal of High Energy Physics} {\bf 2024} (June, 2024).

\bibitem{harrow2013churchsymmetricsubspace}
A.~W. Harrow, {\it The church of the symmetric subspace},  2013.

\bibitem{Akers:2025ahe}
C.~Akers, G.~Bueller, O.~DeWolfe, K.~Higginbotham, J.~Reinking, and R.~Rodriguez, {\it {On observers in holographic maps}},  {\em JHEP} {\bf 05} (2025) 201, [\href{http://arxiv.org/abs/2503.09681}{{\tt arXiv:2503.09681}}].

\bibitem{Bousso:2025fgg}
R.~Bousso and S.~Kaya, {\it {Fundamental Complement of a Gravitating Region}},  \href{http://arxiv.org/abs/2505.15886}{{\tt arXiv:2505.15886}}.

\bibitem{Akers:2021lms}
C.~Akers, S.~Hern\'andez-Cuenca, and P.~Rath, {\it {Quantum Extremal Surfaces and the Holographic Entropy Cone}},  {\em JHEP} {\bf 11} (2021) 177, [\href{http://arxiv.org/abs/2108.07280}{{\tt arXiv:2108.07280}}].

\bibitem{Iliesiu:2024cnh}
L.~V. Iliesiu, A.~Levine, H.~W. Lin, H.~Maxfield, and M.~Mezei, {\it {On the non-perturbative bulk Hilbert space of JT gravity}},  {\em JHEP} {\bf 10} (2024) 220, [\href{http://arxiv.org/abs/2403.08696}{{\tt arXiv:2403.08696}}].

\bibitem{Held:2024qcl}
J.~Held, X.~Liu, D.~Marolf, and Z.~Wang, {\it {Euclidean and complex geometries from real-time computations of gravitational R\'enyi entropies}},  {\em JHEP} {\bf 02} (2025) 136, [\href{http://arxiv.org/abs/2409.17428}{{\tt arXiv:2409.17428}}].

\bibitem{Dong:2023xxe}
X.~Dong, D.~Marolf, and P.~Rath, {\it {Constrained HRT Surfaces and their Entropic Interpretation}},  {\em JHEP} {\bf 02} (2024) 151, [\href{http://arxiv.org/abs/2311.18290}{{\tt arXiv:2311.18290}}].

\bibitem{dong2019flat}
X.~Dong, D.~Harlow, and D.~Marolf, {\it Flat entanglement spectra in fixed-area states of quantum gravity},  {\em Journal of High Energy Physics} {\bf 2019} (2019), no.~10 1--25.

\bibitem{akers2019holographic}
C.~Akers and P.~Rath, {\it Holographic {R}enyi entropy from quantum error correction},  {\em Journal of High Energy Physics} {\bf 2019} (2019), no.~5 1--24.

\bibitem{penington2022replica}
G.~Penington, S.~H. Shenker, D.~Stanford, and Z.~Yang, {\it Replica wormholes and the black hole interior},  {\em Journal of High Energy Physics} {\bf 2022} (2022), no.~3 1--87.

\bibitem{Ivo:2024ill}
V.~Ivo, Y.-Z. Li, and J.~Maldacena, {\it {The no boundary density matrix}},  \href{http://arxiv.org/abs/2409.14218}{{\tt arXiv:2409.14218}}.

\bibitem{Harlow:2018tqv}
D.~Harlow and D.~Jafferis, {\it {The Factorization Problem in Jackiw-Teitelboim Gravity}},  {\em JHEP} {\bf 02} (2020) 177, [\href{http://arxiv.org/abs/1804.01081}{{\tt arXiv:1804.01081}}].

\bibitem{Chua:2023srl}
W.~Z. Chua and T.~Hartman, {\it {Black hole wavefunctions and microcanonical states}},  {\em JHEP} {\bf 06} (2024) 054, [\href{http://arxiv.org/abs/2309.05041}{{\tt arXiv:2309.05041}}].

\bibitem{Jia:2020etj}
H.~F. Jia and M.~Rangamani, {\it {Petz reconstruction in random tensor networks}},  {\em JHEP} {\bf 10} (2020) 050, [\href{http://arxiv.org/abs/2006.12601}{{\tt arXiv:2006.12601}}].

\bibitem{kaya2025}
S.~Kaya, P.~Rath, and K.~Ritchie, {\it To appear}, .

\end{thebibliography}\endgroup
\end{document}